\documentclass[12pt]{article}

\usepackage{scicite}

\usepackage{times}
\usepackage{authblk}
\usepackage{graphicx}
\usepackage{multirow}
\usepackage{siunitx}
\usepackage{rotating}
\usepackage{diagbox}
\usepackage{bm}

\newenvironment{sciabstract}{%
\begin{quote} \bf}
{\end{quote}}

\title{Frequency Ratio Measurements with 18-digit Accuracy Using a Network of Optical Clocks}

\author
{Boulder Atomic Clock Optical Network (BACON) Collaboration$^{1,2}$: Kyle Beloy$^{1}$, Martha I. Bodine$^{1}$, Tobias Bothwell$^{2}$, Samuel M. Brewer$^{1,3}$, Sarah L. Bromley$^{2,4}$, Jwo-Sy Chen$^{1,5,6}$, Jean-Daniel Desch\^enes$^{1,7}$, Scott A. Diddams$^{1,5}$, Robert J. Fasano$^{1,5}$, Tara M. Fortier$^{1\ast}$, Youssef S. Hassan$^{1,5}$, David B. Hume$^{1\ast}$, Dhruv Kedar$^{2}$, Colin J. Kennedy$^{1,2\ast}$, Isaac Khader$^{1}$, Amanda Koepke$^{1\ast}$, David R. Leibrandt$^{1,5}$, Holly Leopardi$^{1,5,8\ast}$, Andrew D. Ludlow$^{1}$, William F. McGrew$^{1,5}$, William R. Milner$^{2}$, Nathan R. Newbury$^{1}$, Daniele Nicolodi$^{1,5}$, Eric Oelker$^{1,2}$, Thomas E. Parker$^{1}$, John M. Robinson$^{2}$, Stefania Romisch$^{1, 9}$, Stefan A. Sch\"{a}ffer$^{1,10}$, Jeffrey A. Sherman$^{1}$, Laura C. Sinclair$^{1\ast}$, Lindsay Sonderhouse$^{2}$, William C. Swann$^{1}$, Jian Yao$^{1}$, Jun Ye$^{1,2,5}$, Xiaogang Zhang$^{1,5,11\ast}$\\
~\\
\normalsize{$^{1}$National Institute of Standards and Technology, Boulder, CO 80305, USA}\\
\normalsize{$^{2}$JILA, University of Colorado, Boulder, CO 80309, USA}\\
\normalsize{$^{3}$Current affiliation: Department of Physics, Colorado State University, Fort Collins, CO 80523, USA}\\
\normalsize{$^{4}$Current affiliation:  Department of Physics, Durham University, Durham, UK}\\
\normalsize{$^{5}$Department of Physics, University of Colorado, Boulder, CO 80309, USA}\\
\normalsize{$^{6}$Current affiliation: IonQ, Inc, College Park, MD 20740, USA}\\
\normalsize{$^{7}$Current affiliation: Consultation OctoSig Inc., Quebec City, Quebec, Canada}\\
\normalsize{$^{8}$Current affiliation: Space Dynamics Laboratory, North Logan, UT 84341, USA}\\
\normalsize{$^{9}$Current affiliation: Northrup Grumman Mission Systems, Woodland Hills, CA 91367, USA}\\
\normalsize{$^{10}$The Niels Bohr Institute, University of Copenhagen, Copenhagen, Denmark}\\
\normalsize{$^{11}$State Key Laboratory of Advanced Optical Communication Systems and Networks, Institute of Quantum Electronics, Department of Electronics, Peking University, Beijing 100871, P. R. China}

\normalsize{$^\ast$To whom correspondence should be addressed; Email: tara.fortier@nist.gov (T.M.F.), david.hume@nist.gov (D.B.H), colin.kennedy@colorado.edu (C.J.K), amanda.koepke@nist.gov (A.K.), holly.leopardi@colorado.edu (H.L.), laura.sinclair@nist.gov (L.C.S.), xiaogang.zhang@nist.gov (X.Z.).}
}

\date{}

\begin{document}

\baselineskip24pt

\maketitle

\begin{sciabstract}
Atomic clocks occupy a unique position in measurement science, exhibiting higher accuracy than any other measurement standard and underpinning six out of seven base units in the SI system. By exploiting higher resonance frequencies, optical atomic clocks now achieve greater stability and lower frequency uncertainty than existing primary standards. Here, we report frequency ratios of the $^{27}$Al$^+$, $^{171}$Yb and $^{87}$Sr optical clocks in Boulder, Colorado, measured across an optical network spanned by both fiber and free-space links. These ratios have been evaluated with measurement uncertainties between $6\times10^{-18}$ and $8\times10^{-18}$, making them the most accurate reported measurements of frequency ratios to date.  This represents a critical step towards redefinition of the SI second and future applications such as relativistic geodesy and tests of fundamental physics.
\end{sciabstract}

\section*{Introduction}
Atoms serve as excellent references for the unit of time because they emit and absorb photons at well-defined resonance frequencies that are largely decoupled from environmental perturbations.  For this reason, in 1967 the General Conference on Weights and Measures adopted a resonance near 9.2 GHz in $^{133}$Cs as the official definition of the SI second.  Currently, this primary reference enables realization of the second with up to 16-digit accuracy limited by the ability to measure and control extremely fine (10 $\mu$Hz level) perturbations of the microwave frequency. However, newly-developed clocks based on optical transitions permit even greater control of fractional frequency perturbations ($\delta\nu/\nu$ for uncertainty $\delta\nu$ of a clock frequency $\nu$) by leveraging resonance frequencies that are 100,000 times higher than those of microwave standards. As a result, compared to primary microwave standards, clocks based on optical transitions have recently demonstrated more than a factor of 100 improvement in both frequency stability and uncertainty due to systematic effects, providing impetus for the eventual redefinition of the SI second \cite{ludlow2015}.

Verifying the performance of these optical clocks requires inter-laboratory frequency comparisons with measurement accuracy approaching the best frequency uncertainties, $\delta\nu/\nu \sim10^{-18}$ \cite{nicholson2015,huntemann2016,mcgrew2018,brewer2019,bothwell2019}. While any absolute measurement of an optical clock frequency, i.e. in Hertz, will be limited by the accuracy of primary $^{133}$Cs standards ($\approx10^{-16}$), frequency ratios between two optical clocks, which are unitless, can be measured with accuracy approaching the uncertainty due to those clocks' systematic effects. With this in mind, guidelines for the redefinition of the SI second recommend the measurement and independent verification of multiple frequency ratios with uncertainties below $5\times10^{-18}$ \cite{riehle2018}. Comparisons within a single laboratory between clocks based on the same atomic species have shown frequency reproducibility approaching \cite{chou2010_1} or even below \cite{mcgrew2018, sanner2018} this level.  In contrast, among the reported frequency ratio measurements of optical clocks based on different atomic species \cite{rosenband2008,godun2014,takamoto2015,nemitz2016,tyumenev2016, ohmae2020}, the lowest measurement uncertainty is $4.6\times10^{-17}$.  In addition, only one ratio, $^{199}$Hg/$^{87}$Sr, has been measured directly at multiple laboratories, with agreement at $2.0\times10^{-16}$ \cite{takamoto2015,tyumenev2016}.

One challenge in verifying clock accuracy internationally is that frequency ratio measurements across intercontinental distances are limited by existing time and frequency transfer methods.  Microwave time and frequency transfer, the current intercontinental state-of-the-art, is limited to measurement stability of approximately $10^{-15}$ at one day of averaging \cite{riehle2017} -- a level optical clocks can achieve in less than 1~s of averaging. Future fiber-based or satellite-based networks (or transportable standards \cite{delehaye2018,grotti2018,takamoto2020}) may offer improvements, but those capabilities are still under development for intercontinental baselines\cite{riehle2017}. In the near term, by comparing frequency ratios between two clocks measured in one regional network with the same ratios measured in other networks, we can compare clocks globally at a level never before achieved. Local and regional networks of optical clocks based on stabilized optical fiber links \cite{foreman2007,takamoto2015,lisdat2016} have been shown to permit frequency ratio measurements between clocks separated by as much as 690 km \cite{lisdat2016} with negligible added noise.

Here, we report the inter-comparison of three optical clocks at the National Institute of Standards and Technology (NIST) and JILA in Boulder, Colorado.  We evaluate their fractional ratio uncertainties to be between 6$\times10^{-18}$ and 8$\times10^{-18}$.  This represents a significant step towards redefinition of the SI second and future applications of optical clock networks.

\section*{The optical clocks}
The frequency ratio measurements presented here involve three optical clocks that are based on the ${{^1\!S_0}\leftrightarrow{^3\!P_0}}$ transitions in $^{27}$Al$^+$, $^{171}$Yb, and $^{87}$Sr.  These atomic transitions feature narrow natural linewidths ($<10$~mHz) at accessible laser wavelengths and are relatively insensitive to perturbation from external electric and magnetic fields \cite{dehmelt1982,hall1989}.  Because these clocks are being actively developed at other metrology laboratories worldwide \cite{ludlow2015}, the ratio measurements reported here can be readily compared to existing and future measurements from independent groups.

The clocks involved in these measurements have fractional uncertainties due to systematic effects evaluated at the $10^{-18}$ level and are representative of the current state-of-the-art for both optical lattice clocks and single-ion clocks. These two platforms are the leading technologies for optical standards, but differ in several ways with respect to frequency stability and systematic effects.  Optical lattice clocks operate with thousands of atoms, which reduces quantum projection noise compared to that of a single ion \cite{itano1993}, leading to intrinsically higher clock stability \cite{mcgrew2018, oelker2019}.  On the other hand, single-ion clocks exhibit lower intrinsic sensitivity to some environmental effects such as the Stark shift due to blackbody radiation \cite{brewer2019}.  In general, systematic shifts to the clock frequency depend on unique properties of each atom, as well as on the design of the experiment and its control systems.  Some of the most important parameters affecting frequency stability and systematic effects are summarized in the supplemental material (Table S1), and details pertaining to the various experimental systems can be found in separate publications (Al$^+$ \cite{brewer2019}, Yb \cite{mcgrew2018}, and Sr \cite{bothwell2019}).

\section*{The optical network}

The three optical clocks in the Boulder network are physically separated by tens to thousands of meters and their frequencies span hundreds of THz. To cover this range of frequencies and locations, the network employs five optical frequency combs (OFCs), which provide coherent synthesis across the optical domain \cite{fortier2019}, and greater than twenty phase-stabilized optical links connecting remote OFCs, lasers and optical clocks.  An overview of this network is given in Fig. 1 (detailed schematic in supplement Fig. S1). Network components were verified separately where possible (dashed lines of Fig. 2).  Additionally, the network contains several redundant elements to test in real-time that measurements are performed with negligible noise penalty or bias.  For instance, two OFCs, a single-branch 180 MHz Er:Fiber \cite{leopardi2017} and an octave-spanning 1 GHz Ti:Sapphire (Ti:S) OFC \cite{fortier2006} were operated throughout the measurement campaign with their simultaneously-recorded ratio values agreeing at the $5\times10^{-19}$ level.

Local to each clock, a laser is pre-stabilized to a high-finesse optical reference cavity, which provides short-term stability and a linewidth sufficiently narrow to probe the clock transition.  The optical clock transition wavelengths near 267 nm, 578 nm and 698 nm are coherently connected to infrared lasers at 1070 nm, 1156 nm and 1542 nm, respectively, which are delivered to the OFCs via optical fibers with active phase noise cancellation \cite{ludlow2015}.  The prestabilized lasers for the Yb and Al$^+$ clocks are near subharmonics of the clock transition frequency while the Sr clock uses another OFC to phase-coherently connect a silicon cavity-stabilized laser at 1542 nm to a separate clock laser at 698 nm, which ultimately probes the clock transition \cite{oelker2019}. The use of a 1542 nm laser is convenient, as this wavelength experiences low losses in fiber and is transmitted from JILA to NIST over the longest fiber link in the network (3.6~km) \cite{foreman2007}.  In all cases, slow feedback ($<1$ Hz bandwidth) based on interrogation of the atomic clock transition with each local oscillator stabilizes the clock laser to the atomic resonance.

In addition to the fiber-based connections between clocks, the network employs a free-space link between JILA and NIST. Comb-based optical two-way time-frequency transfer (O-TWTFT) is used to measure the Yb/Sr ratio across this link.  Laser light stabilized by the Sr and Yb clocks is transmitted via noise-canceled fibers to OFCs located within transceivers at either end of the free-space optical link.  Comb pulses from each transceiver are exchanged across the link. Since the sampled air paths are common to pulse trains traveling in each direction, O-TWTFT allows for frequency transfer at better than $10^{-18}$ instability despite the presence of km-length turbulent air paths \cite{deschenes2016,sinclair2018}.  In contrast to Ref. \cite{sinclair2018}, which was limited to relative phase measurements and optical sources closely centered around 200 THz, this represents the first use of O-TWTFT to compare state-of-the-art optical atomic clocks (see Ref. \cite{bodine2020} for details).

Accurate ratio evaluation requires frequency corrections to account for each clock’s different gravitational redshift with respect to a reference gravitational potential. While the elevations of the two labs at NIST are approximately the same, the JILA lab is lower by approximately 18.6 m. As part of the measurement campaign, a geodetic survey was conducted in collaboration with NOAA. The survey provided precise elevations  and geopotential differences between permanent markers fixed in the clock labs and a common reference marker (Q407) \cite{van_westrum2016}, which has previously been evaluated with respect to the global mean geoid \cite{pavlis2017}.  In addition, the height of the atomic samples relative to the local lab marker was measured. The final relative frequency shifts at the atomic samples with respect to the common reference marker are $965.3(3)\times 10^{-18}$, $961.1(4)\times 10^{-18}$ and $-1085.9(4)\times 10^{-18}$  for the Yb, Al$^+$ and Sr clocks, respectively.

\section*{Results and data analysis}

The clock comparison spanned November 2017 through June 2018 and consisted of 17 measurements of the Al$^{+}$/Yb ratio, 8 measurements of the Al$^{+}$/Sr ratio and 9 measurements of the Yb/Sr ratio, with each measurement averaging all data acquired on a particular day.  Fig. 2 shows an example of the time series data, obtained on March 6, 2017. The total measurement duration during this day was approximately 13 hours,  with the fractional uptime of the Al$^{+}$ ratios reaching 60 \%, and that of the Yb/Sr ratio reaching 88 \%. The overlapping Allan deviation of the ratios was used to evaluate their stability. The instability in the ratio data is observed to follow white frequency noise statistics, $\sigma(\tau) = \sigma_0/\sqrt{\tau}$,  for averaging periods $\tau$ greater than 100 s. A fit of the parameter $\sigma_0$ to the data yields $3.1\times 10^{-16}$ for the Yb/Sr ratio, and $1.3\times 10^{-15}$ for the Al$^{+}$/Yb and Al$^{+}$/Sr ratios.  Uncertainty due to measurement instability on each day was calculated by extrapolating this fit to the duration of the full data set.

Data for each ratio were corrected from their laboratory operational conditions to the perturbation-free ideal at zero electric and magnetic field, absolute zero temperature, and to a common gravitational potential. Fig. 3 shows the measurement results as a fractional offset from their final reported values.  To evaluate the final ratios and their uncertainties, the data collected on each day were treated as independent measurements for two reasons. First, individual days were operationally different from one another, resulting in both measurement stability and some systematic effects that varied day by day (details in Tables S4-S5). Second, the data were sampled unevenly, with dead-time lasting from minutes to months. Under these conditions, standard tools in stability analysis such as the Allan deviation do not accurately capture the statistical uncertainty except in the case of white frequency noise. While data taken within each day exhibit white frequency noise statistics (Fig. 2), the same was not observed when data were concatenated across measurement days (see Fig. S3).

To estimate the final ratio uncertainties, we consider the daily statistical uncertainties, uncertainties due to systematic effects, and the scatter of the measurements of Fig. 3.  As an initial step, the scatter in the data relative to the daily statistical uncertainties can be quantified by the reduced-$\chi^2$: $\chi^2_{\rm red} = 1/(N-1)\times\sum_{i=1}^N (x_i - \bar{x})^2/\sigma_i^2$.  Here $x_i$ is the daily ratio, $\bar{x}$ is the weighted mean with weights $w_i = 1/\sigma_i^2$ and $\sigma_i$ is the daily statistical uncertainty, which includes contributions from the ratio instability and, in the case of the Sr ratios, uncertainty due to the atomic density shift (see Supplemental Materials, including Table S5). The data in Fig. 3 have  $\chi^2_{\rm red}= \{1.5, 0.2, 6.0\}$ for Al$^+$/Yb, Al$^+$/Sr and Yb/Sr, respectively. In the case of the Al$^+$ ratios, with the $\sigma_i$ ranging from $7.3\times10^{-18}$ to $5.4\times10^{-17}$ depending primarily on the duration of the measurement, the $\chi^2_{\rm red}$ values are on the order of one or lower. Conversely, the Yb/Sr measurements had $\sigma_i$ as low as $2.3\times10^{-18}$ in a single day, revealing excess scatter beyond statistical effects alone. We interpret this as evidence for systematic frequency shifts that appear constant on a particular day but vary between days. While we have not identified a specific source of these systematic shifts (see Fig. S4), we describe all known clock systematic effects in the supplemental information and references therein.

One commonly used approach to evaluate the statistical error in the presence of excess scatter is the standard error of the mean inflated by $\sqrt{\chi_{\rm red}^2}$ (Birge ratio \cite{mohr2016}), which is labelled as weighted standard error (WSE) in Table 1. In the case of Al$^+$/Sr with $\chi^2_{\rm red} < 1$, we use the standard error without deflating by $\sqrt{\chi_{\rm red}^2}$. The total uncertainty from the combined data can then be estimated by summing the WSE in quadrature with uncertainties due to systematic effects as shown in Table 1. This approach gives a reasonable first estimate for the uncertainty, but has the drawback that it treats the Al$^+$/Sr ratio differently based on its statistical properties and it does not yield specific information about the excess variability between days.

A more rigorous value for the ratio uncertainty can be assessed using a comprehensive Bayesian model, which treats all of the ratios in a single framework. This model incorporates uncertainty due to the known statistical and systematic effects but also allows for unknown effects that may vary between days \cite{gelman2014,koepke2017}.  The analysis determines the posterior distributions of different model parameters given the observed data. Detailed model parameters for each ratio, including prior and posterior distributions for the mean values and excess variability (see Fig. S5 and Fig. S6) are all outlined in the supplemental material. The main results of this analysis are: 1) agreement of the consensus ratio values with the weighted means from the simplified analysis above, 2) a more conservative estimate of the final ratio uncertainties, as seen in Table 1, and 3) credible intervals for the between-day variability (see supplemental material).

Based on the results of the comprehensive Bayesian model, the final reported ratio values and their 1 $\sigma$ uncertainties are:
\begin{eqnarray}
f_{\rm Al^+}/f_{\rm Yb} &=& \num{2.162887127516663703(13)}\nonumber \\
f_{\rm Al^+}/f_{\rm Sr} &=& \num{2.611701431781463025(21)}\nonumber \\
f_{\rm Yb}/f_{\rm Sr}   &=& \num{1.2075070393433378482(82)}.\nonumber
\end{eqnarray}

Although the final fractional uncertainties are comparable for all ratios, the statistical and systematic effects that contribute to each one are distinct. We summarize those observations here:

\emph{Al$^+$/Yb}: With 17 total measurement days, the value $\chi^2_{\rm red} = 1.5$ is only marginally significant (probability $P = 0.1$ of observing a value higher than this by statistical fluctuations alone). The ratio uncertainty is dominated by measurement instability, primarily due to quantum projection noise of the single Al$^+$ ion. In contrast, the combined uncertainty due to systematic effects of $2.2\times10^{-18}$ has a minor impact on the final ratio uncertainty.

\emph{Al$^+$/Sr}: The uncertainty in this ratio has approximately equal contributions from clock instability and the total uncertainty due to systematic effects of $5.1\times10^{-18}$. While the ratio data appear to be underscattered, the comprehensive Bayesian model still includes a term for between-day fluctuations, which increases the final uncertainty compared to the evaluation based on the standard error.

\emph{Yb/Sr}: The lower instability of the lattice clocks allows for the resolution of systematic effects which vary between measurement days. The largest contribution to the total uncertainty of this ratio is thus the combined uncertainty due to systematic effects of $5.2\times10^{-18}$.  Another significant contribution comes from the between-day variability which is quantified using the comprehensive Bayesian model.

\section*{Comparison between the free-space and terrestrial fiber links}

The network presented here incorporates both fiber and free-space connections between remote clocks. Fig. 4a provides a sketch of the vertical cross section of the network highlighting the free-space link.  Simultaneous measurements of the Yb/Sr ratio using the free-space and fiber links provide a real-time test of the network connections independently of the clocks. The O-TWTFT system was operated for 6 days in conjunction with the rest of the network during the measurement campaign. As an example, the time series of the data from March 6, 2018 is provided in Fig. 4b.  For each day, a daily mean and a statistical uncertainty is computed with the uncertainty determined by evaluating the overlapping Allan deviation as in Fig 2.  Fig. 4d shows a strong correlation between the Yb/Sr ratio as measured via these two optical links, constraining the NIST-JILA link stability and frequency offsets. When restricted to common measurement times the weighted mean of the difference in ratios obtained by the two links is $(-4.5 \pm 6.1)\times10^{-19}$ (see also \cite{bodine2020}).

\section*{Conclusion}

We report the first measurements of frequency ratios with fractional uncertainties below $1\times 10^{-17}$. Figure 5 compares our results with previous frequency ratio measurements. For all ratios, we observe consistency with CIPM (International Committee for Weights and Measures) recommended clock frequency values \cite{riehle2018}.  Among these, only the Yb/Sr ratio has been previously measured via optical comparison; our result is in agreement with the weighted mean of all the previous optical measurements within 1.7 $\sigma$ (where $\sigma$ is the standard error of the weighted mean). In the calculation of this mean and uncertainty we assume no correlation in the measurements contributing to these ratios, an assumption that needs careful consideration as more ratios are measured.  The previous ratios that include Al$^+$ all rely on an optical comparison with the Hg$^+$ optical clock \cite{rosenband2008} combined with an absolute frequency measurement of the Hg$^+$ clock \cite{stalnaker2007}, which strongly correlates these data points. Here, for both Al$^+$/Sr and Al$^+$/Yb, we observe a significant difference from those previous measurements, with discrepancies based on their uncertainties as high as 3 $\sigma$, which we have not been able to explain with any evaluations of the current systems.

Ultimately, the gains in measurement accuracy reported here resulted from improved performance of each clock \cite{mcgrew2018,brewer2019,bothwell2019}. Careful characterization and control of the clock systematic frequency shifts during the campaign permitted fractional uncertainties for each individual clock at or below $5\times10^{-18}$. We exploited redundancy in the optical links and measurement combs to characterize the performance of individual network components, demonstrating their uncertainty to be below that of all the optical clocks.  We also developed a Bayesian framework for rigorously evaluating each ratios' expected value and its uncertainty despite their different statistical properties.  Finally, the two-way optical free-space link between NIST and JILA is the first to be used in a high-accuracy optical clock comparison to date. Via simultaneous measurement with a terrestrial fiber link over the course of six measurement days, we demonstrated agreement in the Yb/Sr ratio over both links at the $6\times 10^{-19}$ level.

The ability to connect clocks at our demonstrated level of accuracy across fiber and free-space optical links represents a critical step towards the future redefinition of the SI second \cite{riehle2015,gill2011}.  In addition, because frequency ratios of optical clocks are defined by, and hence sensitive to, the laws of physics and the fundamental constants, they can be used as sensitive tests of the Standard Model and probes for new physics \cite{safronova2018}.  Examples include testing the predictions of relativity theory \cite{sanner2018} and searching for undiscovered particles that have been proposed to solve problems with the Standard Model such as the nature of dark matter \cite{safronova2018}.  Technological applications for optical clocks are also being pursued including relativistic geodesy \cite{mehlstaubler2018,mcgrew2018,takamoto2020} and very long baseline interferometry \cite{clivati2017}.  With applications like these within reach, the reported measurements represent a significant advancement towards new capabilities in diverse areas of science and technology.

\newpage

\section*{Supplemental Text}

\section{Clock operational details}

The design, operating conditions and evaluation of systematic effects of the three optical clocks involved in these measurements are described in detail in separate publications \cite{mcgrew2018,brewer2019,bothwell2019}.  Table S1 provides an overview of some of the most important parameters.  Below, we describe some additional considerations relevant for the specific operating conditions during the measurement campaign.

\subsection{Al$^+$ single-ion clock}

The Al$^+$ clock is based on quantum-logic spectroscopy of a single Al$^+$ ion co-trapped with a single $^{25}$Mg$^+$ ion for sympathetic cooling and state detection \cite{chou2010}.  Before the clock probe pulse, the ions are cooled close to the ground state for all six normal modes of motion via resolved-sideband Raman transitions on $^{25}$Mg$^+$ \cite{chen2017}.  The clock laser at 267.4 nm is derived from frequency quadrupling a Yb-doped fiber laser, and stabilized to a high-finesse ULE cavity near 534.9 nm.  This laser is stabilized to the atomic resonance via Rabi spectroscopy with a probe duration of 150 ms and a total average cycle time of 300 ms, with the remaining duty cycle used for laser cooling, quantum-logic state detection, optical pumping and auxiliary locks to control micromotion, which is a dominant systematic effect \cite{brewer2019}. Two counter-propagating clock probe beams are used alternately to measure and correct for a 1st-order Doppler shift caused by slow ($\approx10$ nm/s) ion motion in the trap.

The systematic effects detailed in \cite{brewer2019} represent the operating condition of the clock during the last few months of this measurement campaign.  Early in the measurement campaign, we observed two effects that caused additional uncertainty in the clock frequency.

First, a linear Doppler shift was observed as a significant offset ($\approx5\times10^{-17}$) in the line centers measured from the two counterpropagating probe lasers.  This is suppressed in all of the data by averaging the direction-dependent error signals, but two effects can lead to this suppression being imperfect: 1) angular deviations between the two counterpropagating beams and 2) contrast imbalance between the resonances observed with the two beams.  The first effect was minimized beginning in data from February 16, 2018 by coupling the two counterpropagating beams through the same optical fiber.  The second effect was eliminated by introducing a differential servo for the two probe directions, which was implemented beginning with the data from March 28, 2018.  The source of the 1st-order Doppler shift was confirmed as motion of the ion relative to the trap by inserting measurements of RF micromotion into the normal clock sequence.  These measurements showed the ion was moving in the trap electric fields at a velocity consistent with the observed 1st-order Doppler shift and provided an independent constraint on the motion. The most likely cause of this motion is charging and subsequent discharging of the trap electrodes by photoelectrons from the UV cooling lasers, which are applied before the clock probe period. Both clock beams were phase-stabilized close to the ion trap, which eliminates phase shifts of the laser as the source of the offset.

Second, some data were affected by phase slips in the phase-locked loops used for cancelling Doppler noise of the optical fibers \cite{ma1994,newbury2007,foreman2007}.  For the Al$^+$ clock, we continuously monitor the beatnote frequency in these phase-locked loops using frequency counters.  Phase slips were detected as sharp deviations of the beatnote frequency from its expected value. To prevent such phase slips from biasing the measured frequency ratio, we remove ratio data 10 s before and 100 s after any observed outliers in the fiber-noise counters.  To bound the possible impact of these outliers, we compare the frequency measured in a 60 second window (beginning 10 s before the events and ending 50 s after the event) to the frequency measured using the data with these spikes removed and find a difference equal to $5(2)\times10^{-17}$.  We model the effect of these spikes as the impulse response of the clock servo, such that the frequency offset they induce is damped with a time constant equal to the measured servo time constant.  Within this model we find the average offset at the end of the 110 s window to be $3\times10^{-19}$, which we take as an upper bound on the uncertainty of the shift, because it will be further suppressed during the uninterrupted lock period.  Nevertheless, this uncertainty is applied to all of the data which had more than 5\% of the data removed due to glitches.

\subsection{Sr optical lattice clock}

This work employed the SrI clock system located at JILA. In 2013, the uncertainty of the JILA SrI clock due to systematic effects was evaluated at $5\times10^{-17}$ with the JILA SrII clock later being evaluated at $2.1\times10^{-18}$ \cite{nicholson2015}. However, the SrII clock was unavailable for this work. Therefore, the uncertainty evaluation of SrI was improved before and throughout this work. As a result, the uncertainty detailed in \cite{bothwell2019} was not fully realized by the end of the data campaign.  There are four differences between the evaluation detailed in Ref. \cite{bothwell2019} and that demonstrated here.

First, in Ref. \cite{bothwell2019} the density shift was carefully controlled for day-to-day reproducibility with an uncertainty of $4\times10^{-19}$.  However, here the atomic density shift was evaluated on each comparison day, yielding a mean density shift coefficient ranging between $-1.71\times10^{-17}$ and $-3.57\times10^{-17}$ per 1000 atoms with each determination having measurement-limited statistical uncertainty between $1.3\times10^{-18}$ and $3.2\times10^{-18}$ per 1000 atoms. Corrections for the density shift are made point-by-point using the measured atom number on each experimental cycle. The average shift and associated uncertainty applied to each point on a given day is listed in Table S5. The average atom number varies between days ranging from 1000 to 2700. The uncertainty of these frequency corrections are statistically limited by each day's evaluation of the density shift coefficient and as such are added in quadrature with the statistical uncertainty for each day's ratio measurement. The total contribution of the Sr density shift uncertainty to the systematic error budget in Table 1 is $1.0 \times 10^{-18}$ for both the Yb/Sr and Al$^{+}$/Sr ratios.

Second, the lattice light shift uncertainty is higher than that reported in Ref. \cite{bothwell2019}. In this work, two different lattice wavelengths are used, one for all the data up to the last day and then -- after replacing the 532 nm pump laser for the Ti:S which supplies 813 nm trapping light -- a different wavelength for the last comparison day. For the first wavelength, six independent evaluations of the lattice light shift were undertaken over the course of four months producing an uncertainty of $3.7 \times 10^{-18}$ at the nominal operational trap depth. Uncertainty in the trap depth determination produces an additional uncertainty ranging from $8.0 \times 10^{-19}$ to $1.9 \times 10^{-18}$ and the effective hyperpolarizability coefficient, $\beta^{*}$, contributes an additional uncertainty of $8.9 \times 10^{-19}$. For the second wavelength -- used only on the final comparison day -- the light shift evaluation taken on the same day contributes an uncertainty of $6.4 \times 10^{-18}$.

Third, the quoted blackbody radiation uncertainty in Ref. \cite{bothwell2019} has been increased compared to the ratio measurements reported here due to aging of both thermistor calibrations and resistance measurement calibrations. The dominant contribution to this uncertainty is the aging of the resistance measurement calibration. Aside from these calibration uncertainties, the thermal environment control detailed in Ref. \cite{bothwell2019} is identical to the one used in this work with an average temperature difference between clock comparison data and that in Ref. \cite{bothwell2019} of -5 mK. During the clock comparison data campaign, the standard deviation of the model temperature on any given day was typically higher than that observed in Ref. \cite{bothwell2019}. To account for this effect, we include an increased temperature uncertainty which varies from a minimum of 900 $\mu$K to a maximum of 12 mK. As a result of the aforementioned effects, the uncertainty in the determination of the blackbody radiation shift is increased from the $2 \times 10^{-19}$ level of Ref. \cite{bothwell2019} to a range which varies day-by-day from $3.6 \times 10^{-19}$ to $8.9 \times 10^{-19}$.

Finally, Ref. \cite{bothwell2019} utilizes an improved fiber phase noise cancellation scheme developed for the optimization of stability transfer of cavity-stabilized probe laser light to the atoms \cite{oelker2019}. As an added benefit, this scheme dramatically reduces shifts due to the clock AOM phase chirp because these phase transients are either entirely in-loop, or occur when the laser frequency is at large detuning where the atomic sensitivity function is zero. However, this improved scheme was not yet implemented in the work presented here and therefore a shift uncertainty of $1.2 \times 10^{-18}$ is included.

\subsection{Yb optical lattice clock}

During each comparison day, the Yb optical lattice clock was operated at the nominal conditions described in reference \cite{mcgrew2018}. Prior to each comparison, operational parameters were carefully measured to confirm that the systematic shifts and uncertainty were consistent with reference \cite{mcgrew2018}.

Recently, an independent experimental effort \cite{nemitz2019} measured the magnetic dipole/electric quadrupole polarizability in Yb to be larger than earlier theoretical calculations \cite{brown2017,katori2015,safronovanodate}. While additional effort will be required to resolve this discrepancy, we note the larger polarizability would imply an additional lattice light shift of approximately $3\times10^{-18}$, which is below the uncertainty of the ratios reported here.

During the comparison, the status of the Yb optical lattice clock is determined by the combined efforts of a digital acquisition system and a human operator. If the clock laser became unlocked from the atomic transition, a human operator was required to recover the lock and verify normal system operation. After such a relock, the subsequent two minutes of data are flagged as bad data for removal, to conservatively account for the reacquisition process.

To verify the systematic effects of the Yb optical lattice clocks at NIST, two similar systems,  Yb-1 and Yb-2, were developed. For all of the optical comparisons, except for March 23, 2018, the Yb-1 clock system was used. On March 23, 2018, the Yb-2 clock system, run under identical conditions to Yb-1, was used for optical frequency comparisons. The frequency consistency of the Yb-1 and Yb-2 systems has been evaluated as $(-7\pm9)\times 10^{-19}$ in reference \cite{mcgrew2018}, and thus does not contribute a significant uncertainty to the optical ratio.

We note that all the optical frequency ratio data contributing to the measurement described here was collected simultaneously to the data used to determine Yb clock absolute frequency reported in reference \cite{mcgrew2019}.

\section{Measurement overview}

The three optical atomic clocks were simultaneously measured using two independent optical frequency combs (OFCs) developed at NIST: a single branch Er:fiber laser \cite{leopardi2017} and an octave-spanning Ti:S laser \cite{fortier2006}. The two OFCs were located in the same lab on independent, floating optical tables with largely independent acoustic enclosures and thermal environments.  The performance of these two OFCs has been previously evaluated, with offsets of $(0.1\pm1.7)\times10^{-20}$, and stabilities in the mid 10$^{-20}$ level, at averaging times of 10,000 s. Measurements from the two OFCs agree at the $5\times 10^{-19}$ level, limited by differential out-of-loop optical paths and by the uncertainties of the counters used for these measurements.

Both OFCs are fully stabilized. The carrier-envelope offset frequency $f_0$ is measured with an f-2f non-linear interferometer and stabilized to a fixed frequency derived from the H-maser. The repetition rate $f_{rep}$ is stabilized by phase-locking the heterodyne beat note between a single optical mode of the comb and the Yb clock laser. The optical heterodyne beat notes, which measure the frequency difference between the optical atomic clock and the nearest OFC mode, are recorded using $\Lambda$-type frequency counters with a 1 s gate time, except for those on February 28, 2018 which were taken with a 10 s gate time. In addition to the heterodyne beat notes between the clocks and the OFC, $f_0$ and $f_{rep}$ are also measured and recorded using frequency counters. All labs included in this measurement used a common hydrogen maser signal as a reference for all synthesizers and frequency counters.

\section{Ratio calculations}

The OFCs do not measure the transition frequency of these atomic clocks directly, instead they measure the frequency of atom-stabilized clock lasers in the infrared. Both the Er:Fiber and Ti:S OFCs measure the Al$^+$ clock laser at 1070 nm (282 THz) and the Yb lattice clock laser at 1156 nm (259 THz). The Er:fiber OFC measures the Sr lattice clock laser at 1542 nm (193 THz), while the Ti:S OFC measures the frequency doubled 1542-nm clock laser at 771 nm (389 THz) because 1542 nm is outside its optical bandwidth.

For each clock, $k\in\{\rm Al^+,Yb,Sr\}$, the clock laser frequency measured at the comb, $f_{{\rm comb},k}$, is given by,
\begin{equation}\label{eq:combeqn}
f_{{\rm comb},k} = n_{k} f_{\rm rep}+f_0 + f_{{\rm b},k},
\end{equation}
where $n_{k}$ is the comb mode number against which the beatnote $f_{{\rm b},k}$ was measured.  Note that these numbers will, in general, be different for each comb, but we are omitting subscripts that identify the comb here for simplicity.  The signs of $f_0$, $f_{\rm b,k}$, and the value of $n_{k}$ are confirmed daily using the CIPM recommended atomic frequencies.  The frequency of the clock transition, $f_{k}$ is then given by,
\begin{equation}
\label{eq:fk}
f_{\rm k} = m_{k} f_{{\rm comb},k} + f_{{\rm offset},k} - f_{{\rm shift},k},
\end{equation}
where $m_{k}$ is a rational multiplier and $f_{{\rm offset},k}$ is a frequency offset, which depends on the network components between each clock and the two combs.   The frequency $f_{{\rm shift},k}$ includes systematic shifts that are recorded (but not corrected) in real time as well as other calibrated frequency shifts particular to each clock.  The final frequency ratio for clock $k$ and clock $l$ is $R_{k,l} = f_{\rm k}/f_{\rm l}$.

\subsection{Counter offsets}

$\Lambda$-type frequency counters were used to measure all the optical heterodyne beat notes for the measurements presented here. To evaluate the performance of these counters, the 10 MHz maser signal used as a common reference was split and sent to the input and reference ports of each counter. The fractional uncertainties in the counted frequencies due to counter offsets were estimated as $(f_{\rm counter} - 10\,{\rm MHz})/10\,{\rm MHz}$ and were applied to all directly-measured optical beatnote frequencies involved in the ratio calculations.  The exceptions to this were the $f_0$ and $f_{b,{\rm Yb}}$ beat signals, which were tightly locked to known values, and $f_{\rm rep}$, which was calculated based on published frequencies as described below.  The fractional uncertainties in the ratios due to counter offsets on each day ranged from $2\times10^{-20}$ to $1\times10^{-18}$ and were included in the total network uncertainty ($\sigma_N$).  This value (main text, Table 1) is the weighted mean of the daily counter offset for each ratio using ratio statistical uncertainties to compute the weights.

\subsubsection{Maser offsets}

In Eqs. 1 and 2, all frequencies are referenced to the same maser with the exception of $f_{{\rm shift}, k}$, which are Hz-level and computed based on independent calibrations.  To determine the maser offset in real time we compare the maser-referenced OFC repetition rate, $f_{\rm rep, maser}$ to its optically referenced value, $f_{\rm rep, opt}$, based on the published absolute frequency of the Yb-lattice clock \cite{riehle2017} and Eqs. 1 and 2.  Since $f_{\rm rep, opt}$ is based on an accurate optical frequency, the effects of counter and maser offsets are suppressed by orders of magnitude compared to the directly measured, $f_{\rm rep, maser}$.

This procedure gives a calibration of the offset of the reference maser:
\begin{equation}\label{maseroffset}
    \epsilon_{\rm maser}= \frac{f_{\rm rep, opt}- f_{\rm rep, maser}}{f_{\rm rep,opt}}.
\end{equation}
Once the fractional offset of the reference maser is determined, any signals generated by a maser-referenced synthesizer including those that are used to phase lock $f_0$ and $f_{b,{\rm Yb}}$ as well as any AOM frequencies present in the optical network, can be corrected by multiplying each term by $(1+\epsilon_{\rm maser})$.  A plot of these offsets, as calculated from the ratio data, is given in Fig. S2 and compared with the maser offsets determined by calibration with UTC(NIST) and UTC.  In the final ratio evaluation we use the optically-referenced repetition rate $f_{\rm rep, opt}$ and correct all synthesized frequencies using the calibrated maser offset.

\subsubsection{Fiber link offsets}

A 3.6 km fiber optic link located under the city of Boulder connects JILA (which houses the Sr lattice clock) and NIST. Two optical fibers are used; one carries 1542 nm light referenced to the Sr clock to NIST for measurement, the other carries a NIST maser signal to JILA via a modulated laser diode \cite{foreman2007,campbell2008}.

To evaluate the performance of the fiber link, the two optical fibers were connected at NIST so that a round trip measurement could be made. By comparing the frequency of the ``local'' Sr clock laser with the ``remote'' clock laser that had traversed the round-trip fiber, we evaluated the instability and systematic offset of this fiber-noise-cancelled (FNC) link. The measurement was done over a two day period, resulting in an offset of $(-2\pm5)\times10^{-19}$.  We take $5\times10^{-19}$ as the ratio uncertainty due to the fiber link.

Shorter FNC links between labs may also introduce additional noise and offsets. For example, light from the Yb clock is transferred to the comb lab via a single 100 m long FNC link which is then sent into free space to be split between the two OFCs. The Er:fiber comb is heterodyned with the Yb light from the FNC link, while some of the light is launched into another 20 m FNC link to be measured by the Ti:S comb. For the Al$^+$ clock, two independent 75-m long fiber links are used to send light to the OFCs. In both of these setups short differential optical paths, where small offsets could be introduced, are present for the light measured by the two OFCs.

To determine if any offsets are introduced by these fiber-noise-cancelled links, a free space heterodyne beat between the two independent fiber links for the Al$^+$ clock was measured using a software-defined radio (SDR) frequency counter.  From this an offset of $(0.0 \pm 1.2)\times10^{-19}$ was measured, where the uncertainty is determined by the modified Allan deviation of the phase difference data. Based on this, we include an estimate of the offsets caused by FNC links of $1.2\times10^{-19}$ in the Network ($\sigma_N$) uncertainty in Table 1.

\section{Analysis}

\subsection{Data filtering}

Initial filtering of ratio data is performed on the beatnote of each clock with the OFC by iteratively discarding points greater than 5 $\sigma$ away from the mean. This identifies and discards cycle slips in the optical frequency network which appear as discrete glitches in the counted frequencies.  The final data for the frequency comparison uses only the common uptime between both of the clocks.  Most clock unlocks are flagged automatically by the experimental control systems.  After discarding these data, another filter is applied to identify unlock events that are not automatically flagged. In this process, data are binned with 20 s and 40 s averaging times and 5 $\sigma$ outliers from the median are removed.  Although some outliers were detected in this process, we find that their exclusion shifts the daily measured mean negligibly compared to the ratio uncertainties.

\subsection{Stability}

For each day's measurement, the stability is evaluated by fitting the model ${\sigma(\tau) = \sigma_{0}/\sqrt{\tau}}$ to the overlapping Allan deviation of the ratio data. For this fit we use averaging intervals ranging from 100 s through $T/6$ where $T$ is the total measurement duration.  The lower bound of 100 s is several times longer than the servo time constant for all of the atomic clocks. Allan deviation results for $\tau > T/6$ are excluded from fitting due to their low statistical power. In addition, averaging intervals separated by factors of two were chosen to minimize the effect of correlations in the Allan deviation points \cite{schlossberger2019}.  Based on the white-noise model, the daily statistical uncertainty due to clock stability is determined by the fit extrapolated to the total length of the data set, $T$.

Both measurement OFCs were used simultaneously, but depending on the measurement day, one or the other OFC would have a slightly shorter data set due to phase slips and unlocks. Since the disagreement between OFCs for all ratios was less than $1\times10^{-19}$, the OFC with the longest measurement record was used for a given day's measurement, such that the total data set includes data from both the Er:fiber and Ti:S OFC systems.

A plot of the long-term stability of the ratios is given in Fig. S3.  This is generated by binning all of the measurement data into 10 s intervals, then concatenating multiple days of data together, ignoring dead-time, and computing the overlapping Allan deviation.  Care should be taken when interpreting Fig. S3 because data were acquired in short segments over many months.  However, similar to what can be seen in the $\chi^2$ analysis in the main text, quantum projection noise dominates the instability of the Al$^+$ ratios for time-scales from 100 s to about 30,000 s.  For the Yb/Sr ratio, with resolution approaching $1\times10^{-18}$, significant effects from fluctuating systematic shifts appear as a departure from $1/\sqrt{\tau}$ scaling at longer averaging times.  This appears at a level consistent with the combined uncertainty due to systematic effects.

\subsection{Correlation analysis}

Since many of the measurements described here were taken with three clocks running simultaneously, a significant linear relationship between the daily measurements of two ratios could identify the source of between-day variability.  To test this, we use errors-in-variables regression \cite{fuller-1987, carroll-2006} to estimate the slope of a line fit through data from all ratio pairs (Fig. S4) while accounting for known statistical uncertainties.  Statistical fluctuations arising from the overlapping uptime of the clock frequency that is shared between the x- and y-axes is perfectly correlated, so uncertainty from this is removed from the total statistical uncertainties of the ratios. Specifically, the total statistical variance for each ratio, $\sigma^2_i$, is separated into independent contributions from the non-common-mode clock ($B$) and the overlapping and non-overlapping portions of the clock which is common to both axes ($A$): $\sigma^2_i = \sigma^2_{A,{\rm over}} + \sigma^2_{A, \rm{non}}+ \sigma^2_{B}$. From the measured ratio stability on each day the instability of the overlapping portion of clock B is estimated by, $s_A/\sqrt{T}\times\sqrt{t/T}$, where $s_A$ is the single clock stability, $T$ is the daily ratio data length, and $t$ is the duration of the overlapping data between the two ratios. This is subtracted from the total uncertainty, yielding an estimate of the uncorrelated variance $\sigma^2_{A,{\rm non}} + \sigma^2_B$.  These error bars allow us to treat the X and Y fluctuations independently. The fitted lines and 95 \% confidence intervals, calculated using a parametric bootstrap approach, are plotted in Fig. S4. The confidence intervals for the three slope parameters each contain zero meaning that the relatively small number of measurements do not provide sufficient statistical power to resolve any significant linear relationships between ratios.

\subsection{Ratio uncertainties estimated from $\chi^2$ statistics}

Analysis of the ratio values and their total uncertainties based on the day-by-day measurements was complicated by two main factors.  First, some operating conditions varied over the course of the measurement campaign, such that both the clock stabilities and some systematic shifts had to be evaluated each day.  This affected the uncertainty assigned to each day and hence the relative weight of that data point when evaluating the ratio mean.  It also affected uncertainty due to systematic shifts, such that we had to allow for frequency shifts that were not identical, but still correlated from day to day.  Second, the statistical properties of the data, in terms of its scatter relative to the evaluated uncertainty, varied significantly between the three ratios.  While noise in the Al$^+$ ratios was predominantly white frequency noise due to higher quantum projection noise of the single ion, noise in the Yb/Sr ratio at long averaging times was dominated by effects that varied from day to day.

A first estimate of the ratio statistical uncertainty was carried out based on a $\chi^2$ analysis of the daily ratios, $x_i$, and their statistical uncertainties, $\sigma_i$, as described in the main text.  The weighted mean is given by ${\bar{x} = \sum_i w_i x_i/(\sum_i w_i)}$, with weights ${w_i = 1/\sigma_i^2}$.  The total statistical uncertainty can be estimated by the weighted standard error, ${WSE = \sqrt{\chi^{2}_{\rm red}}\times(\sum_{i}\frac1{\sigma_i^2})^{-1/2}}$, where $\chi^2_{red} = 1/(N-1)\times\sum_{i=1}^N (x_i - \mu)^2/\sigma_i^2$ and $N$ is the total number of measurements. The weighted standard deviation represents the mean day-to-day variation in the observed ratio and is given by $WSD = \sqrt{N}\times WSE$.  For our data, WSD ${ = \{1.7, 0.6, 1.1\}\times10^{-17}}$ for Al$^+$/Yb, Al$^+$/Sr and Yb/Sr, respectively.  Since $\chi^2_{\rm red} = 0.2$ for the Al$^+$/Sr ratio, to avoid underestimating the total uncertainty, we use $\sigma_{stat} = (\sum_{i}\frac1{\sigma_i^2})^{-1/2}$ in the initial estimate for the statistical uncertainty of the Al$^+$/Sr ratio.  Because the systematic shifts for Al$^+$ and Sr varied across the measurement days, we estimated their total uncertainty contribution in Table 1 as a weighted mean of the daily values with the same weights given above.

\subsection{Comprehensive Bayesian model}

While metrics such as the weighted standard deviation have been commonly used for individual frequency ratio analyses based on the noise character of the measurements involved, we developed a comprehensive Bayesian approach for the computation of all three frequency ratio values and their associated uncertainties.  This allowed us to treat all three frequency ratios in an equivalent way based on our best knowledge of their noise properties.  Additionally, such an approach provides not only an estimate of the ratio uncertainty but also an estimate of the uncertainties of other model outputs such as an additional between-day variability not included in the clocks evaluations.

We describe ratio-specific parameters in Sec. 4.5.2, but each can be summarized using the following framework for a given ratio measurement. We define $\mu$ as the true unknown value of the ratio, expressed in all cases below as an offset from the ratio of frequency values currently recommended by CIPM \cite{riehle2018}. Systematic effects cause us to measure instead a ratio $\eta$, shifted from $\mu$. Formally, systematic effects that have constant magnitude across the measurement campaign, such as those due to the geopotentials or the comparison network, yield a measured ratio
$$\eta\vert\mu,\sigma_C \sim N(\mu,\sigma_C^2),$$
where the vertical bar can be read as ``given'', ``$\sim$'' means ``is distributed as'', and $N(m,v)$ denotes a normal distribution with mean $m$ and variance $v$. Here $\sigma_C^2$ is the quadrature sum of the uncertainties due to static systematic effects (reported in Table 1). This $\eta$ will be the same for every individual ratio measurement taken over the entire measurement campaign.

Of course, on any given day of ratio measurements we do not directly measure even this shifted value. Rather, the daily ratio measurement $x_i$ for day $i =1,\ldots,I$ will be affected by various random and systematic effects that vary from day to day. Some systematic effects may be correlated from day to day but scaled by different values corresponding to the operating conditions. For example, uncertainty in the sensitivity of a clock frequency to an electromagnetic field may be scaled by a known field at which the clock was operated on a particular day. We make the conservative assumption that these systematic effects are fully correlated across all days of measurement and broadly address these correlations by defining systematic effects $\alpha \sim N(0,1)$ and $\beta \sim N(0,1)$, one for each clock, where $\alpha$ and $\beta$ are fixed across the measurement campaign. These systematic effects shift the measured daily ratios from $\eta$. The magnitude of the effect is scaled by a daily vectors $\mathbf{a}$ and $\mathbf{b}$ for each clock. Each component $a_i$ or $b_i$ defines the magnitude of the clock uncertainty for day $i$.

The daily ratio measurement $x_i$ will also be affected by measurement instability and any other statistical effects that are not correlated from day to day, such as the density shift for the Sr clock. The variance of these known statistical effects is $\sigma_{i}^2$. Finally, in addition we allow for any potential systematic shifts or statistical fluctuations that are not accounted for in the evaluations of the clocks. This between-day variability, $\xi^2$, models any excess variation between measurements taken on different days. Combining the known correlated systematic effects, the known statistical effects, and any unknown excess between-day variability, the ratio measurements $x_i$ can be modeled as
$$ x_i\vert\eta,\alpha,\beta,\mathbf{a},\mathbf{b},\xi,\boldsymbol{\mathbf{\sigma}} \sim N(\eta+a_i\alpha+b_i\beta,\sigma^2_i+\xi^2).$$

We use this framework to develop a specific model for the measurements of each ratio, then use a Bayesian approach to obtain a comprehensive estimate of the true ratio $\mu$ and its uncertainty. The results of this analysis are given in the final row of Table 1. In addition, the Bayesian analysis yields a value and uncertainty for any excess between-day variability through $\xi^2$.

Alternatively, we can think of this as an additive error model, where the measurement for each day is a sum of the true ratio, $\mu$, and individual random effects
$$ x_{i} = \mu + \lambda_{i} + \epsilon_{i} +a_{i} \alpha +b_{i} \beta + C.$$
The random effects that vary by day, $\lambda_{i}$, are assumed to be independent and identically distributed $N(0,\xi^2)$ over $i$, where the $\xi$ is unknown. The $\epsilon_i$ represents a different random effect due to day, and we assume it has a normal distribution with mean zero and with variance equal to $\sigma_i^2$, which incorporates all known daily varying uncertainties. The $a_i \alpha$ and $b_i \beta$ are as defined above, and $C$ is a systematic effect that does not vary by day which has a normal distribution with mean zero and variance $\sigma_C^2$. This is equivalent to the previously defined hierarchical model.

\subsubsection{Bayesian analysis}

To estimate the parameters of these models, we use a Bayesian approach \cite{gelman2014} similar to that detailed in \cite{koepke2017}. In a Bayesian analysis, we are interested in the posterior probability of our parameters given our data \(p(\boldsymbol{\mathbf{\theta}}\vert\mathbf{y})\), where \(\boldsymbol{\mathbf{\theta}}\) is a vector of unknown parameters and \(\mathbf{y}\) is a vector of the observed data. Using the notation above, \(\boldsymbol{\mathbf{\theta}}=(\mu,\eta,\alpha,\beta,\xi)\) and \(\mathbf{y} = (\mathbf{x},\boldsymbol{\mathbf{\sigma}},\mathbf{a,b}, \sigma_{C})\).  From Bayes' rule, we know
\[p(\boldsymbol{\mathbf{\theta}}\vert\mathbf{y}) \propto {p(\boldsymbol{\mathbf{\theta}})} \times {p(\mathbf{y}\vert\boldsymbol{\mathbf{\theta}})}.\]
Here \(p(\boldsymbol{\mathbf{\theta}})\) is referred to as the prior distribution, and \({p(\mathbf{y}\vert\boldsymbol{\mathbf{\theta}})}\) is the likelihood. The prior distribution \(p(\boldsymbol{\mathbf{\theta}})\) is the product of individual prior distributions for each of the unknown parameters.

For all three ratio models, we assume \textit{a priori} \(\mu \times 10^{18} \sim N(0,10^{5})\), which is a relatively flat distribution that conveys little prior knowledge about the value of the true ratio. For the between-day variability \(\xi\) ($\times 10^{18}$), we assume a half-Cauchy prior distribution centered at zero and with a large scale, which again conveys little prior information. The half-Cauchy distribution is a commonly used prior for between-day variability, since it restricts the parameter to be greater than zero and has long tails which allow for larger values. For this analysis, the scale is set to 1000 ($1\times10^{-15}$ in terms of fractional offset in the ratio) to make the prior distribution diffuse, shown in Fig. S5. We tested the sensitivity of the results to this choice of prior distribution and saw little affect for reasonably diffuse prior distribution options.

With these weakly informative prior distributions, and the likelihood as defined from the above equations, we use Markov chain Monte Carlo (MCMC) to sample from the posterior distribution, implemented using the \texttt{R} package \texttt{R2jags} \cite{su2015}. For computational reasons we multiply all ratios and uncertainties by $10^{18}$ to avoid issues with finite precision math. We use $10^9$ total iterations for the MCMC with $10^7$ burn-in iterations. This means we discard the first $10^7$ samples, where the Markov chain is still converging to the posterior distribution. We thin the resulting chains to reduce autocorrelation in the samples by saving only every 250th sample. The results are thus based on 3960000 samples from the posterior distribution. Generating these samples takes less than five hours on a standard desktop PC. We check for convergence of the Markov chains by visually inspecting trace plots of the posterior samples to make sure the chains are mixing well, as seen in Fig. S7, and using the convergence diagnostic suggested by Ref. \cite{geweke1992}.

We plot the resulting posterior distributions for our parameters of interest focusing on \(\mu\) and \(\xi\). These are shown for the three ratios in Fig. S6. To obtain estimates and uncertainties for these parameters, we calculate the mean and standard deviation of the posterior samples, and the posterior credible interval comes from the 95\% quantiles of this sample. These are reported in Table S2.  As in the main text, we note that the lower of the 95\% credible range values of $\xi$ given in Table S2 are consistent with the total uncertainties due to systematic effects ($2.2\times10^{-18}$, $5.0\times10^{-18}$ and $5.2\times10^{-18}$ for Al$^+$/Yb, Al$^+$/Sr and Yb/Sr respectively).  This is expected as these uncertainties set the scale for the expected uncontrolled variations in the ratio.  This agreement with the lower bound is an indication that use of the comprehensive Bayesian analysis with a random effects model is an appropriate choice for handling the scatter seen in the daily measurements.

\subsubsection{Ratio-specific models for Bayesian analysis}


For the Al$^+$/Yb ratio, we assume
\[\eta\vert\mu,\sigma_N,\sigma_G \sim N(\mu,\sigma_G^2+\sigma_N^2).\]
Here \(\mu\) is the true unknown value of the Al$^+$/Yb ratio. Due to systematic effects, we do not actually measure \(\mu\) but instead measure some random variable \(\eta\) with mean \(\mu\) and variance equal to the quadrature sum of the uncertainties due to the systematic effects. For the Al$^+$/Yb ratio, the relevant systematic effects (and their uncertainties) are for the network (\(\sigma_N=0.3\times10^{-18}\)) and for geopotential (\(\sigma_G=0.2\times10^{-18}\)). These static systematic effects do not vary by day, affecting every measurement in the same way.

The daily ratio measurement \(x_i\) for day \(i =1,\ldots,I\) can be modeled as
\[ x_i\vert\eta,\alpha,\beta,\mathbf{a},\mathbf{b},\xi,\boldsymbol{\mathbf{\sigma}} \sim N(\eta+a_i\alpha+ b_i\beta,\sigma^2_i+\xi^2).\]
The \(a_i \alpha\) term captures the systematic shifts for the Al$^+$ clock and the \(b_i \beta\) term captures the same for the Yb clock. We assume \(\alpha \sim N(0,1)\) and \(\beta \sim N(0,1)\), and \(a_i\) and \(b_i\) are reported in Table S4. The \(\sigma_{i}\) incorporate all known, daily varying uncertainties, which, for the Al$^+$/Yb ratio, is just the stability, $\sigma_{stab,i}$. These values corresponds to the left error bars in Fig. 3 and are reported in Table S5. The between-day variability \(\xi^2\) models any excess variation between measurements taken on different days.


Similarly, for the Al$^+$/Sr ratio, we assume
\[\eta\vert\mu,\sigma_N,\sigma_G \sim N(\mu,\sigma_G^2+\sigma_N^2).\]
where \(\mu\) is now the true, unknown Al$^+$/Sr ratio. Note that, while we have reused symbols in the notation for this model, they are different from the parameters in the Al$^+$/Yb model. Due to systematic effects from geopotential and the network, we don't observe \(\mu\) but instead measure \(\eta\) which has a normal distribution centered at \(\mu\) with variance \(\sigma_G^2+\sigma_N^2\). For this ratio, \(\sigma_G=0.4\times10^{-18}\) and \(\sigma_N=0.5\times10^{-18}\).

We assume the observed Al$^+$/Sr ratio for day \(j\)
\[ x_j\vert\eta,\alpha,\gamma,\mathbf{a},\mathbf{c},\xi,\boldsymbol{\mathbf{\sigma}} \sim N(\eta+a_j\alpha+c_j\gamma,\sigma^2_j+\xi^2).\]
Again \(\alpha \sim N(0,1)\), \(\gamma \sim N(0,1)\), the \(a_j \alpha\) term captures the systematic shifts for the Al$^+$ clock, and the \(c_j \gamma\) term captures the same for the Sr clock. Table S4 shows \(a_j\) and \(c_j\). The \(\sigma_{j}^2\) incorporate all known, daily varying uncertainties. For the Al$^+$/Sr ratio that includes both the stability and a daily varying systematic effect for the density shift for Sr, shown in Fig. 3 and given in Table S5, so $\sigma_{j}^2 = \sigma_{stab,j}^2+\sigma_{SrDen,j}^2$. We assume these effects are not correlated from day to day. The \(\xi^2\) is the between-day variability.


For the Yb/Sr ratio, we assume
\[\eta\vert\mu,\sigma_N,\sigma_G \sim N(\mu,\sigma_G^2+\sigma_N^2),\]
where \(\mu\) is the true, unknown Yb/Sr ratio, \(\sigma_G=0.4\times10^{-18}\) and \(\sigma_N=0.5\times10^{-18}\). We model the Yb/Sr ratio for day \(k=1,\ldots,K\) as
\[ x_k\vert\eta,\beta,\gamma,\mathbf{b},\mathbf{c},\xi,\boldsymbol{\mathbf{\sigma}} \sim N(\eta+b_k\beta+c_k\gamma,\sigma^2_k+\xi^2).\]
Again \(\beta \sim N(0,1)\), \(\gamma \sim N(0,1)\), the \(b_k \beta\) term captures the systematic shifts for the Yb clock, and the \(c_j \gamma\) term captures the same for the Sr clock. Table S4 shows \(b_k\) and \(c_k\). The \(\sigma_{k}^2\) includes both the stability and a daily varying systematic effect for the density shift for Sr, shown in Fig. 3 given in Table S5. Specifically, $\sigma_{k}^2 = \sigma_{stab,k}^2+\sigma_{SrDen,k}^2$. The \(\xi^2\) is the between-day variability.

\newpage




\section*{Acknowledgments}
The authors thank Ethan Clements, Aaron Hankin, Shimon Kolkowitz, Jacob Scott, and Blaza Toman for technical contributions and Antonio Possolo, Christian Sanner and Andrew Wilson for careful reading of the manuscript. \textbf{Funding:} This work is supported by the National Institute of Standards and Technology, Defense Advanced Research Projects Agency, Air Force Office for Scientific Research, National Science Foundation (NSF Grant No. PHY- 1734006) and Office of Naval Research (ONR Grant No. N00014-18-1-2634). \textbf{Author contributions} All authors contributed to the design of the experiment, collection of data, and revision of the manuscript.  During the measurement campaign, Al$^+$ clock operation was conducted by S.M.B., J.S.C., D.B.H., and D.R.L.; Sr clock operation was conducted by T.B., S.L.B., D.K., C.J.K., W.R.M., E.O., J.M.R., L.S., and J.Ye; Yb clock operation was conducted by K.B., R.J.F., Y.S.H., A.D.L., W.F.M., D.N., S.A.S., and X.Z.; comb metrology lab operation was conducted by  T.M.F., H.L.; maser operation and comparison with UTC-NIST was conducted by T.E.P., S.R., J.A.S., and J.Yao; O-TWTFT system operation including the free-space link was conducted by M.I.B., J.D.D., S.A.D., I.K., N.R.N., L.C.S., and W.C.S.; network interconnections excluding the free-space link were maintained by H.L., W.R.M., E.O., and J.M.R. Ratio data analysis and preparation of the manuscript were performed by T.M.F., D.B.H, D.K., C.J.K, A.K., H.L., L.C.S., X.Z. \textbf{Competing interests:} The authors have no competing financial interests. \textbf{Data and materials availability:} All data and supporting materials are available upon reasonable request.

\newpage

\begin{figure}[t!]
\includegraphics[width=1.0\columnwidth]{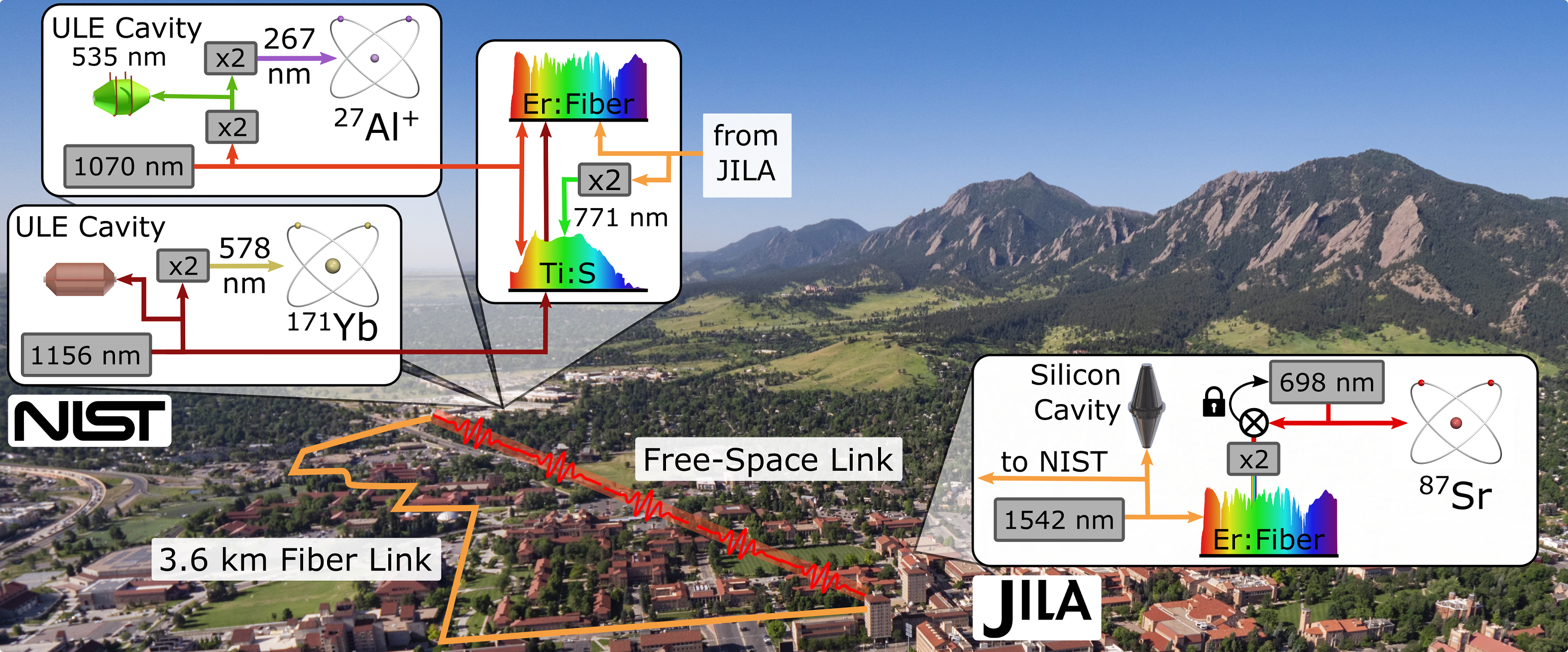}
\caption{System overview of the Boulder Atomic Clock Optical Network (BACON).  Two optical clocks (Al$^+$ and Yb) and two optical frequency combs (Er:fiber and Ti:S) are located on the NIST campus, while the Sr lattice clock is located at JILA.  Stabilized light from the Sr lattice clock is transferred to NIST via a 3.6 km optical fiber link for measurement with the Er:fiber and Ti:S optical frequency combs. The Yb and Sr lattice clocks are also compared using optical two-way time-frequency transfer via a 1.5 km free-space optical link between the two institutions (see also Fig. 4).  Solid colored lines represent noise-canceled optical fiber links. ULE: ultra-low expansion glass.  Photo credit: Glenn Asakawa/University of Colorado Boulder.}
\end{figure}

\newpage

\begin{figure}[t!]
\centering
\includegraphics[width=0.6\columnwidth]{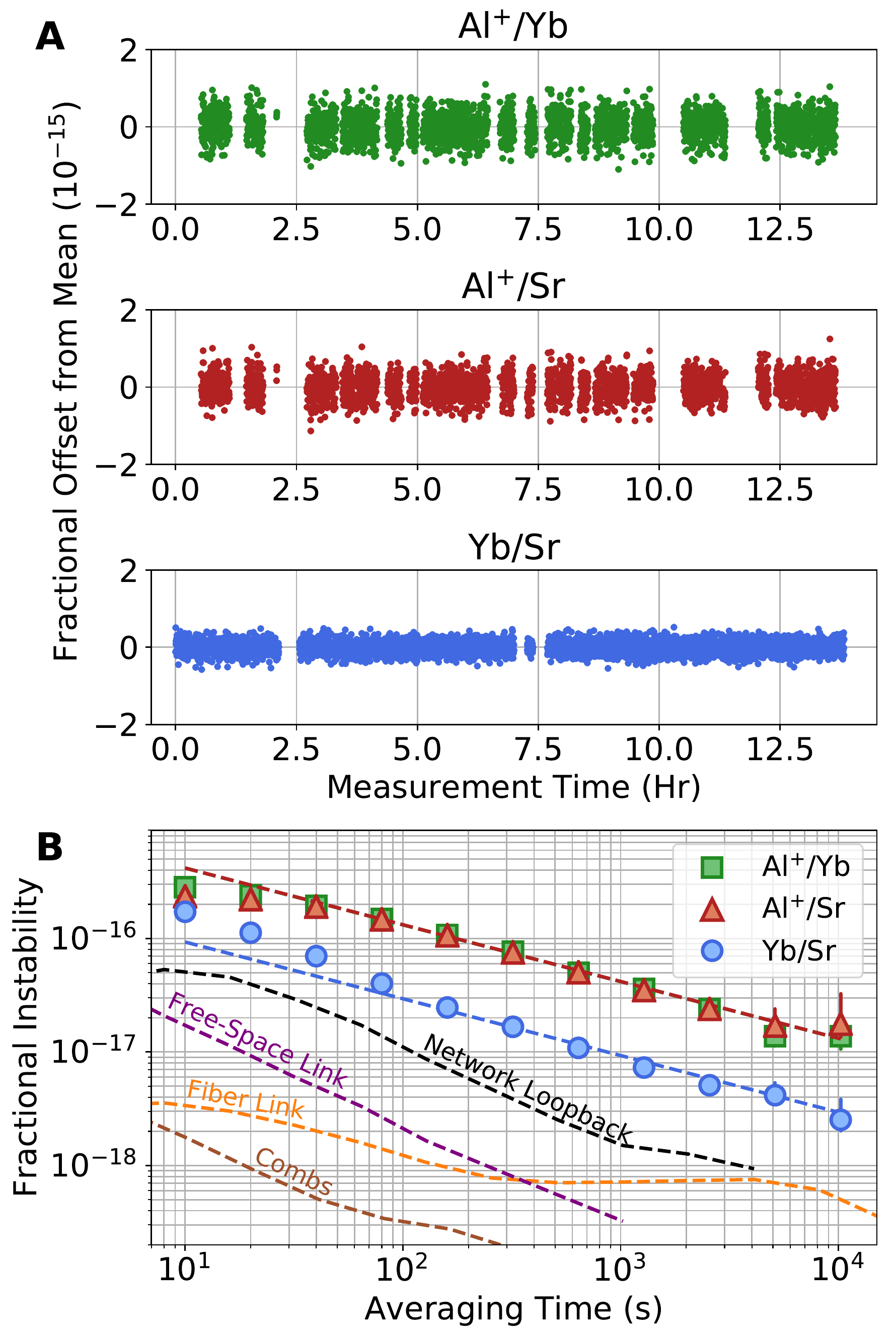}
\caption{A representative sample of data from the measurement campaign (March 6, 2018). (a) Fractional offsets of each ratio from their mean value; each point represents an average over a 10 s interval. (b) Fractional instability in the ratio measurements (Al$^+$/Yb - green squares, Al$^+$/Sr - red triangles, Yb/Sr - blue circles) calculated from the data in (a) using the overlapping Allan deviation.  Matching dashed lines are weighted fits, using a white frequency noise model, to the ratio data beginning at a 100 s averaging time, which yield $3.1\times10^{-16}/\sqrt{\tau}$ for Yb/Sr, and $1.3\times10^{-15}/\sqrt{\tau}$ for both Al$^+$/Sr and Al$^+$/Yb.  Extrapolation of the fit to the full data length was used to determine the statistical uncertainty for each day, assuming white frequency noise. The stability of different network components are also shown, including optical frequency combs (brown, \cite{leopardi2017}), the fiber link between JILA and NIST (orange),  free-space O-TWTFT system (purple, \cite{sinclair2018}) and the stability of a loop-back test (black), which measures the total round trip frequency noise including the free-space link and multiple fiber links between several labs in JILA and NIST.}
\end{figure}

\newpage

\begin{figure}[t!]
\includegraphics[width = 1.0\columnwidth]{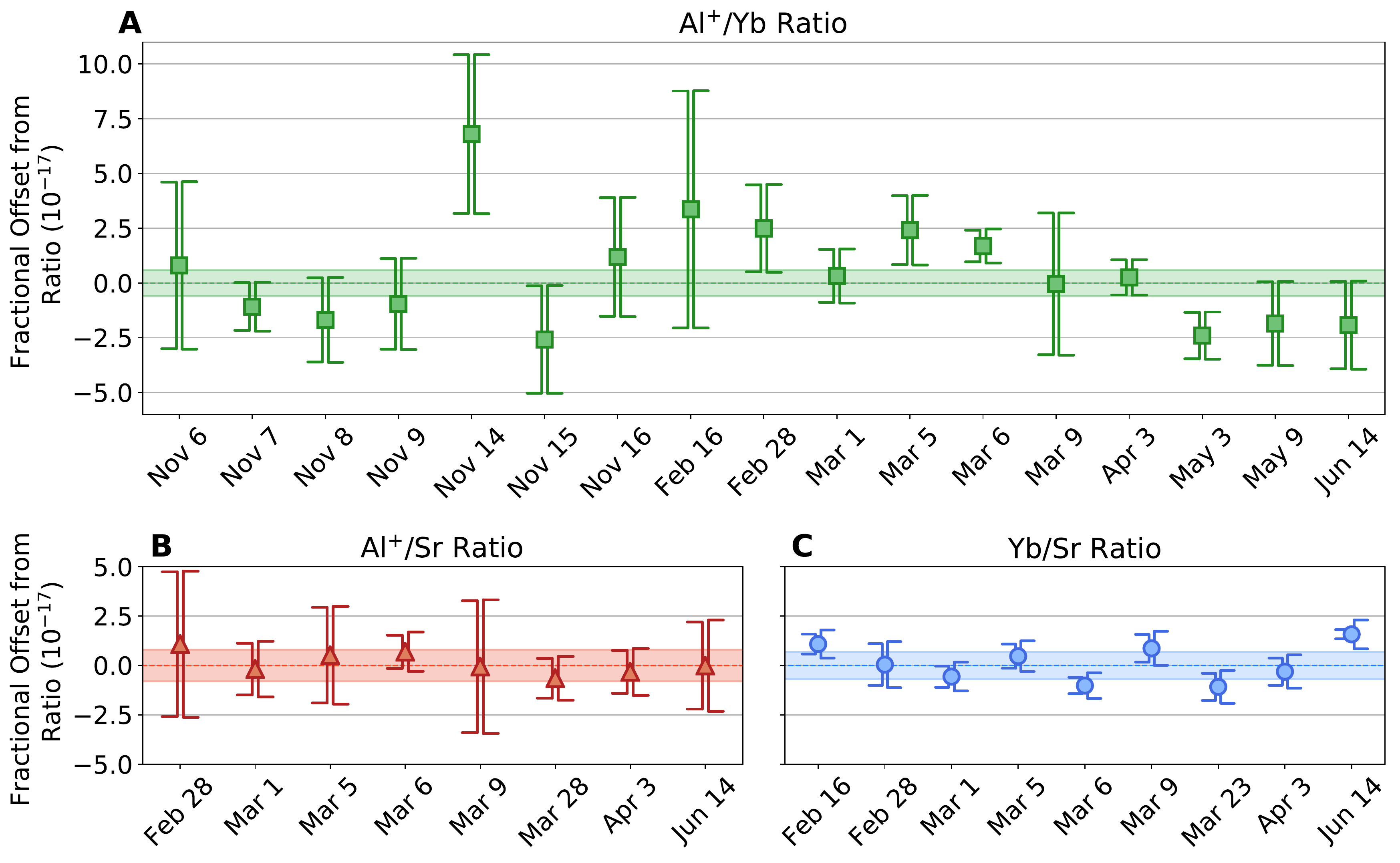}
\caption{Frequency ratio measurements taken from November 2017 to June 2018. The values for each measurement day are displayed as a fractional offset from each ratio's final reported value as computed using the Bayesian analysis. Error bars to the left side of each data point represent the daily statistical uncertainty, $\sigma_i$. For the Al$^{+}$/Yb ratio, statistical error bars, $\sigma_i = \sigma_{\mathrm{stab}, i}$, are evaluated entirely from fits to the overlapping Allan deviation (OADEV) of the sampled ratio data.  For the two ratios that include Sr, $\sigma_i^2 = \sigma_{\mathrm{stab}, i}^2+\sigma_{\mathrm{SrDen},i}^2$, where $\sigma_{\mathrm{SrDen}, i}$ is statistical uncertainty determined by a fit to the OADEV of the Sr density shift data on each day.  Error bars on the right represent the quadrature sum of $\sigma_i$ with all uncertainties due to systematic effects. Lightly shaded regions correspond to the final 1~$\sigma$ uncertainty of each ratio: $5.9\times 10^{-18}$, $8.0\times 10^{-18}$, $6.8\times 10^{-18}$, for Al$^{+}$/Yb, Al$^{+}$/Sr, and Yb/Sr, respectively.}
\end{figure}

\newpage

\begin{figure}[t!]
\includegraphics[width=1.0\columnwidth]{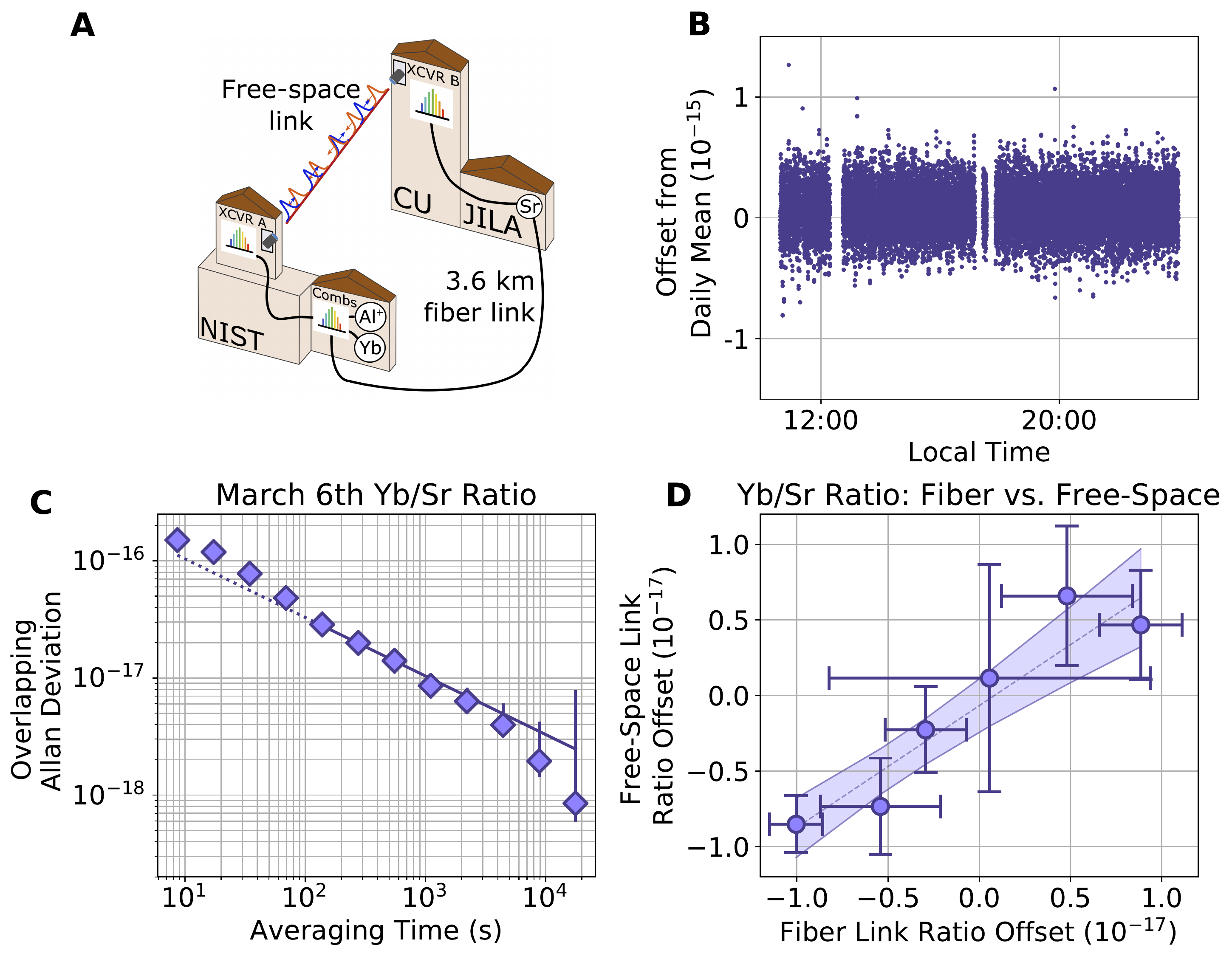}
\caption{(a) Sketch highlighting the comparison of the 1.5~km separated Yb and Sr clocks using optical two-way time frequency transfer (O-TWTFT) which enables a frequency ratio measurement independent of noise from turbulence or free-space path length variations. Along with a fully self-referenced comb \cite{sinclair2015}, the O-TWTFT transceivers (XCVR A and B) each contain an optical heterodyne module, a signal processor, and a free-space optical terminal \cite{bodine2020}. Transceivers are located in a 5$^{th}$ floor laboratory at NIST and in an 11$^{th}$ floor conference room at the University of Colorado (CU). Black lines indicate fiber-noise cancelled links. (b) Time series measurement of the Yb/Sr ratio offset recorded on March 6, 2018 and (c) the corresponding fractional instability characterized by the overlapping Allan deviation (purple diamonds) along with a fit to a white noise model for averaging times beyond 100 s. (d) Free-space vs fiber link daily ratio measurements with a common offset removed demonstrating the expected positive slope. Error bars are derived solely from daily ratio measurement precision excluding the Sr density shift. The two measurement means agree to $6\times 10^{-19}$. The high correlation (Pearson coefficient of 0.89) reflects the dominant sources of noise are the clocks themselves, which are common to both measurements. Shaded region is a 1 $\sigma$ confidence interval for an orthogonal distance regression fit.}
\end{figure}
\newpage

\begin{figure}[t!]
\centering
\includegraphics[width = 0.8\columnwidth]{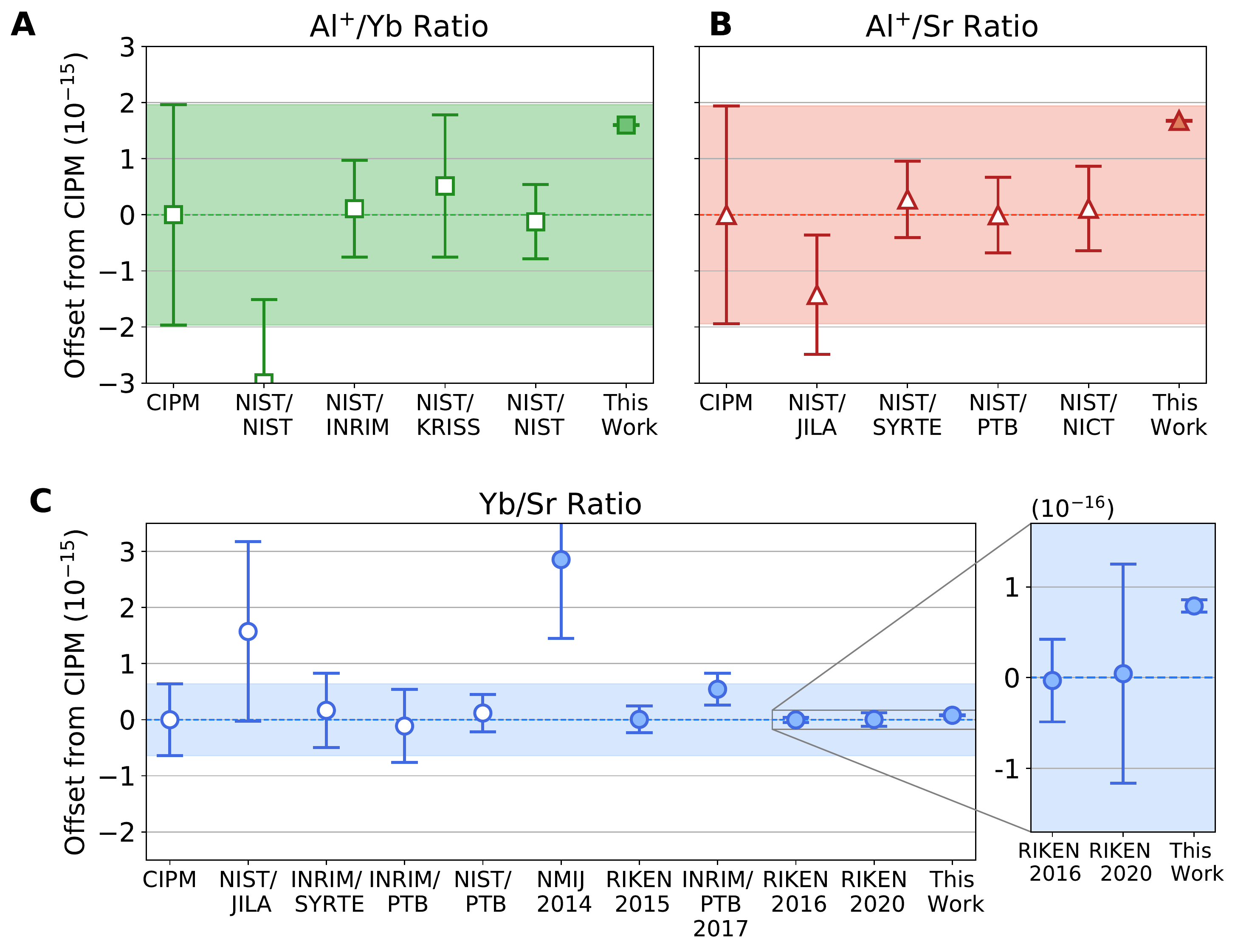}
\caption{Comparison of measured frequency ratios with previous results. All values are reported as fractional offsets from the ratios of current CIPM recommended frequencies (Al$^+$: $\num{1121 015 393 207 857.3}$ Hz, Yb: $\num{518 295 836 590 863.6}$ Hz, and Sr: $\num{429 228 004 229 873.0}$ Hz, with fractional uncertainties of $1.9\times10^{-15}$,  $5\times10^{-16}$, and $4\times10^{-16}$, respectively) \cite{riehle2018}. The shaded region in each plot indicates the standard uncertainty of the CIPM ratio, determined from the quadrature sum of the absolute frequency uncertainties. Ratios based on absolute frequency measurements against $^{133}$Cs are presented with open symbols, whereas direct optical frequency ratios are presented with shaded symbols. For Yb/Sr, a subset of the most accurate absolute frequency measurements is displayed \cite{campbell2008, lemke2009,pizzocaro2017,lodewyck2016,grebing2016,mcgrew2019} in addition to all direct optical frequency ratios \cite{akamatsu2014,takamoto2015,nemitz2016,grotti2018}. For the ratios involving Al$^{+}$, there are no other direct optical measurements against Sr or Yb, so all prior data are based on absolute frequency measurements \cite{stalnaker2007,rosenband2008,hachisu2017,kim2017}. The inset in (c) re-scales the last three points on the plot including this work and the measurement reported by RIKEN in 2016 and 2020 \cite{nemitz2016, ohmae2020}}
\end{figure}

\newpage

\begin{table}[t!]
\centering
\begin{tabular}{|c|r|c|c|c|}
\cline{3-5}
\multicolumn{2}{l|}{} & Al$^+$/Yb $\left(10^{-18}\right)$ & Al$^+$/Sr $\left(10^{-18}\right)$ & Yb/Sr $\left(10^{-18}\right)$ \\
\hline
\multirow{5}{*}{Sys.} & Sr & - & 4.8 & 5.0 \\
& Yb & 1.4 & - & 1.4 \\
& Al$^+$ & 1.7 & 1.5 & - \\
& Network ($\sigma_N$) & 0.3 & 0.5 & 0.5 \\
& Geopotential ($\sigma_G$) & 0.2 & 0.4 & 0.4 \\
\hline
Stat. & WSE $ = \sqrt{\chi^{2}_{\rm red}}\Big(\sum_{i}\frac{1}{\sigma_i^{2}}\Big)^{-1/2}$ & $4.3 \: \left(\chi^{2}_{\rm red} = 1.5\right)$ & $4.8^{\dagger} \: \left(\chi^{2}_{\rm red} = 0.2\right)$ & $3.7 \: \left(\chi^{2}_{\rm red} = 6.0\right)$ \\
\hline
\multicolumn{2}{|r|}{\textbf{Quadrature Sum (sys. and stat.)}} & \textbf{4.8} & \textbf{7.0} & \textbf{6.4} \\
\hline\hline
\multicolumn{2}{|r|}{\textbf{Comprehensive Bayesian model}} & \textbf{5.9} & \textbf{8.0} & \textbf{6.8} \\
\hline
\end{tabular}
\caption{\label{tab:errors} Fractional ratio uncertainties. Contributions to measurement uncertainty due to systematic effects (sys.) from each clock were evaluated separately for all ratios as a weighted mean of the daily uncertainties. The weights are determined by the statistical weight of each measurement day in the computation of the ratio value. Additive uncertainty from the optical network and geopotential were determined by a loopback measurement and a geodetic survey, respectively. We compare an estimate of the total uncertainty calculated from the weighted standard error (WSE, stat.) and uncertainty in systematic effects to the results of the comprehensive Bayesian model, which returns a distribution of ratio values whose standard deviation is given in the last line. ($\dagger$) To avoid underestimating uncertainty in the under-scattered Al$^{+}$/Sr ratio ($\chi^{2}_{\rm red} = 0.2$) the standard error of the mean, ignoring $\chi^2_{\rm red}$, is used.}
\end{table}

\newpage
\renewcommand\thefigure{S\arabic{figure}}
\setcounter{figure}{0}

\begin{sidewaysfigure}[t!]
\centering
\includegraphics[width = 1.0\columnwidth]{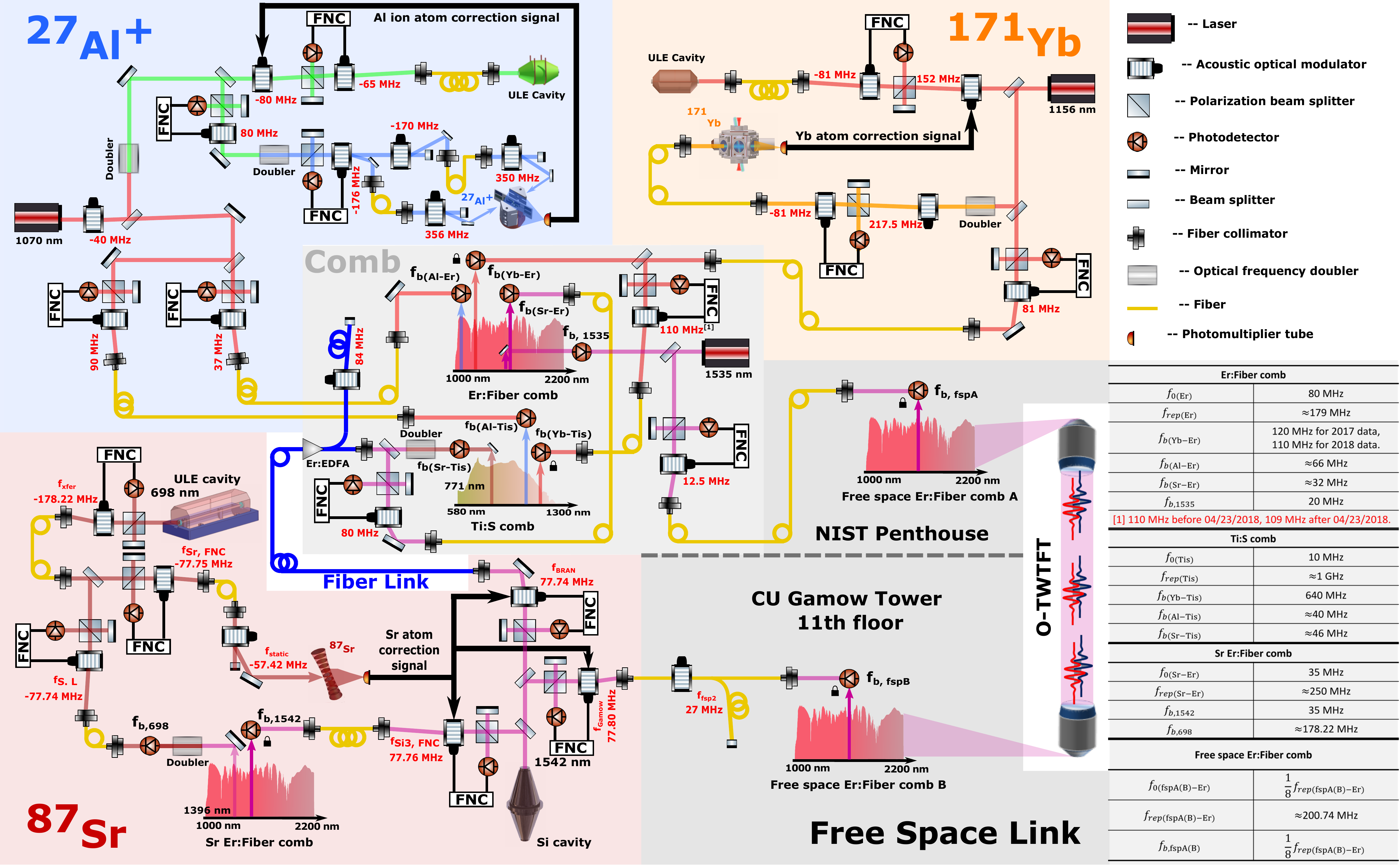}
\caption{Diagram of the Boulder Atomic Clock Optical Network (BACON). The Al$^{+}$ ion optical clock (blue shaded area), the Yb optical lattice clock (orange shaded area), the Er:fiber comb (light gray shaded area) and the Ti:S comb (light gray shaded area) are located in Building 81 of NIST. The Sr optical lattice clock (pink shaded area) is located in the basement of JILA. The free space link (gray shaded area) includes two parts (split by the dashed line): one part is located in the NIST Building 1 penthouse and another part is located at the 11th floor of the Gamow tower at the University of Colorado. The Er:fiber comb and the Ti:S comb at NIST are locked to the Yb optical clock. The Er:fiber comb in the Sr optical clock lab is locked to the Si cavity and is used to transfer the Si cavity stability to the Sr clock laser. Free space Er:fiber combs are locked to the Yb optical clock through the Er:fiber comb at NIST and the Sr optical clock through the Si cavity, respectively. All AOM frequencies in the  network are referenced to a hydrogen maser at NIST, which is transferred to JILA through an optical fiber link. The frequency shifts of optical clocks due to calibrated systematic frequency shifts are added to the optical frequency ratio calculations in post-processing. FNC: Fiber-noise canceled.}
\end{sidewaysfigure}

\newpage

\begin{figure}[t!]
\centering
\includegraphics[scale = 0.5]{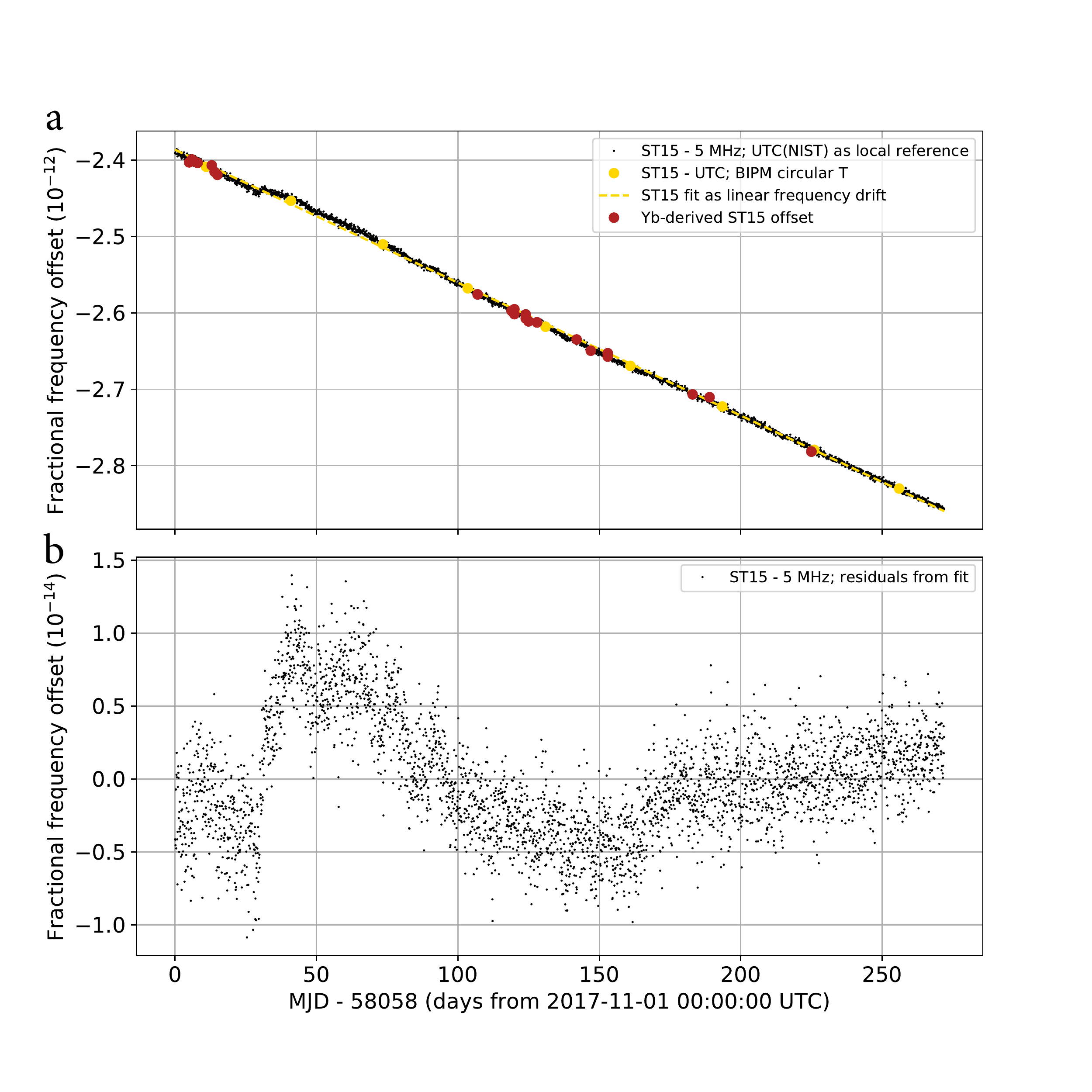}
\caption{Fractional frequency offset of maser ST15 from nominal frequency.  (a) Values used in the data analysis (red points) are compared to values derived from a comparison with UTC(NIST) (black points) and UTC (yellow) as extracted from the BIPM circular T publication.  (b) Residuals show deviations from linear frequency drift across the entire measurement campaign constrained to $\pm2\times10^{-14}$, which leads to negligible fractional uncertainties ($<10^{-19}$) in the ratio values. MJD is the Modified Julian Date.}
\end{figure}

\newpage

\begin{figure}[t!]
\centering
\includegraphics[scale = 0.75]{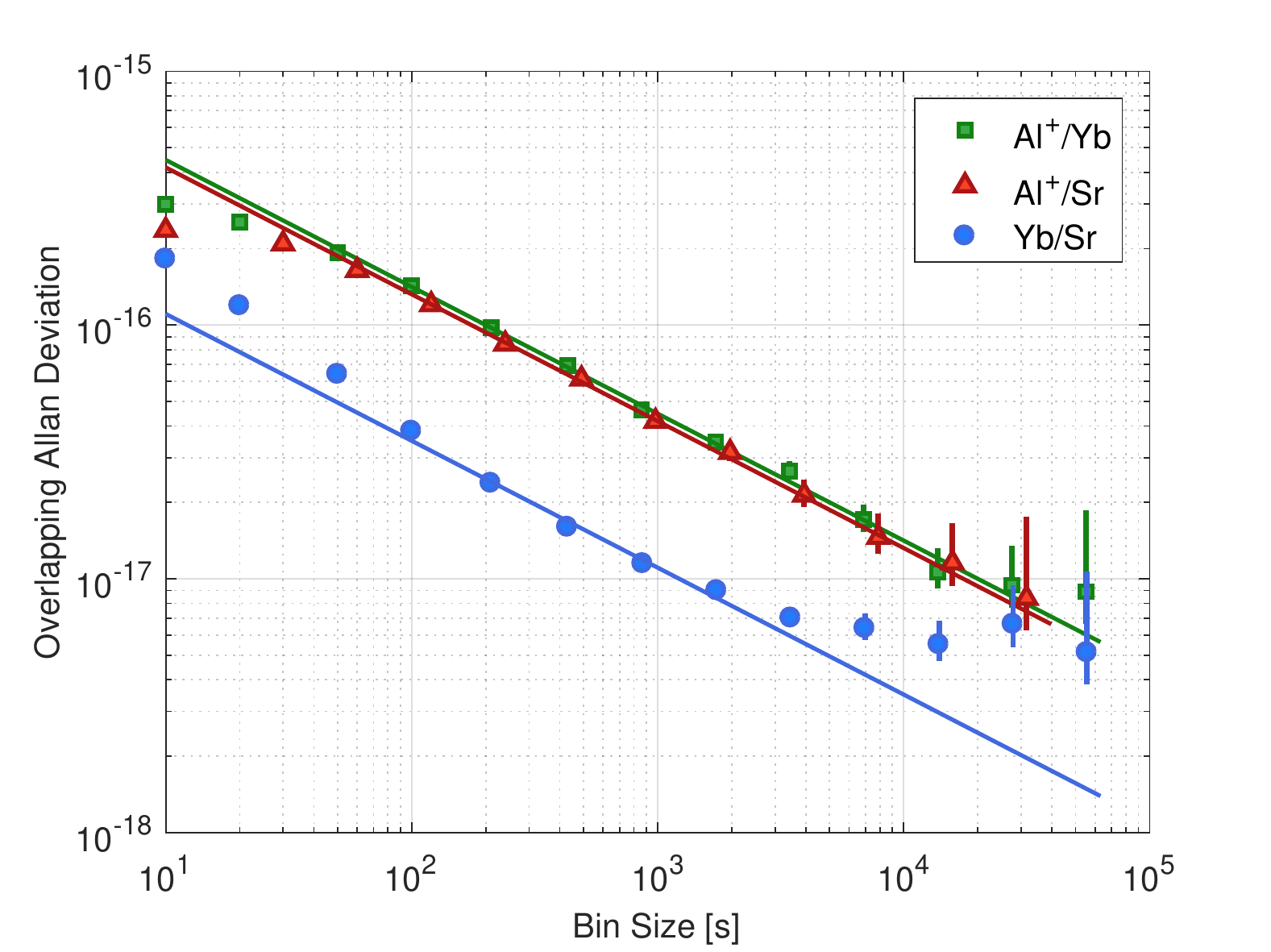}
\caption{Long term stability of concatenated ratio data as characterized by the overlapping Allan deviation.  The data plotted include all measurements that were taken at the nominal operating conditions over the course of the measurement campaign with total measurement durations: 165,240 s, 94,760 s and 167,140 s for Al$^+$/Yb, Al$^+$/Sr and Yb/Sr, respectively.  For the Al$^+$/Yb ratio, in addition to the data that contributed to the final ratio, we include two extra days of data (February 27, 2018 and March 2, 2018) that were used in the evaluation of the Al$^+$ second-order Zeeman shift as described in \cite{brewer2019}. While acquired in very similar experimental conditions, these data sets are excluded from frequency ratio estimate to avoid statistical correlation between the systematic shift evaluation and the frequency ratio measurement.  Since the data were taken in short segments over many months, the time series is dominated by periods of dead-time such that the noise spectrum cannot be identified unambiguously.  Fits (solid lines) use a white frequency noise model: $\sigma_y(\tau) = \sigma_{1s}/\tau^{1/2}$, where $\sigma_{1s}$ is the extrapolated 1-s instability.  These include all data beyond 100 s averaging time with weights equal to the number of bins contributing to each point.}
\end{figure}

\newpage

\begin{figure}[t!]
\centering
\includegraphics[width = 1.0\columnwidth]{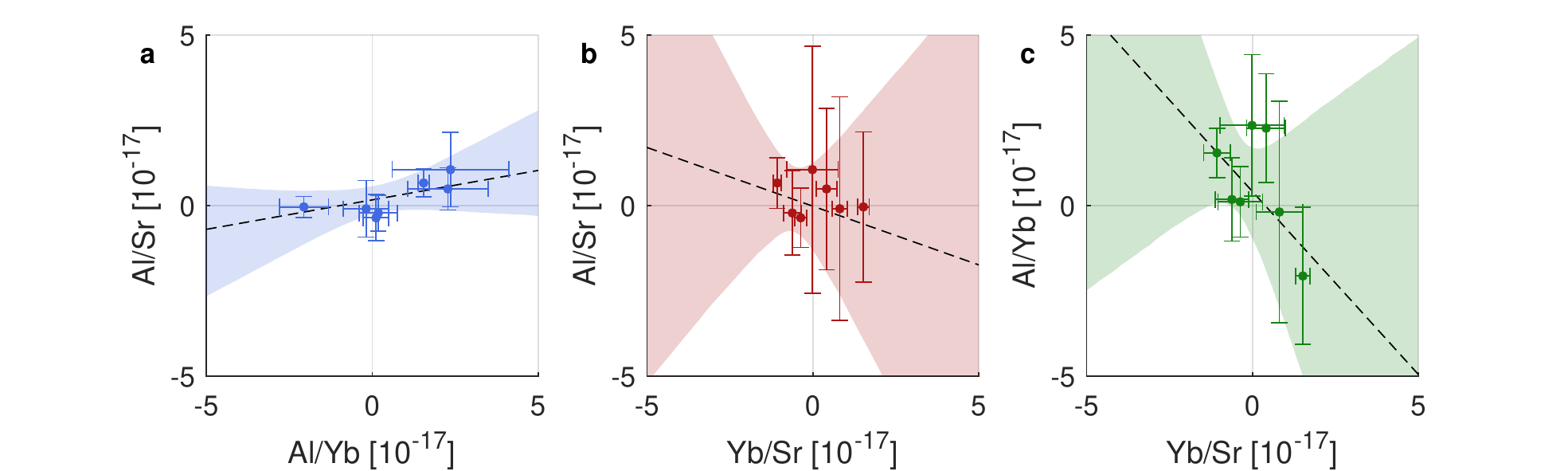}
\caption{Daily measured ratio values (offset by their means) for all ratio pairs with error bars corrected to account for correlations between the x- and y-uncertainties as described in the text.   All days with simultaneous ratio measurements from each pair were plotted. There is no statistically significant linear relationship between these ratios indicating that the present clock data, with only 7 overlapping days of data, are not precise enough to identify the source of daily fluctuations. The slopes and 95 \% confidence intervals for the three plots are: a) 0.17 (-0.08, 0.53), b) -0.34 (-1.90, 1.19), and c) -1.07 (-3.09, 0.72)}
\end{figure}

\newpage

\begin{figure}[t!]
\centering
\includegraphics[width = 0.5\columnwidth]{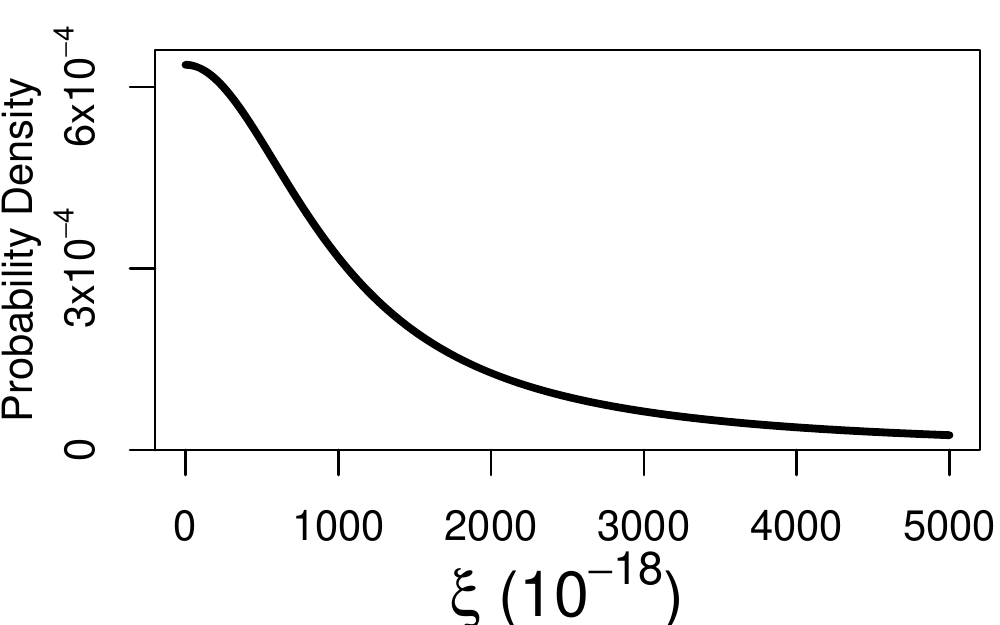}
\caption{Diffuse prior distribution for $\xi$.}
\end{figure}

\newpage

\begin{figure}[t!]
\centering
  \includegraphics[width = 0.75\columnwidth]{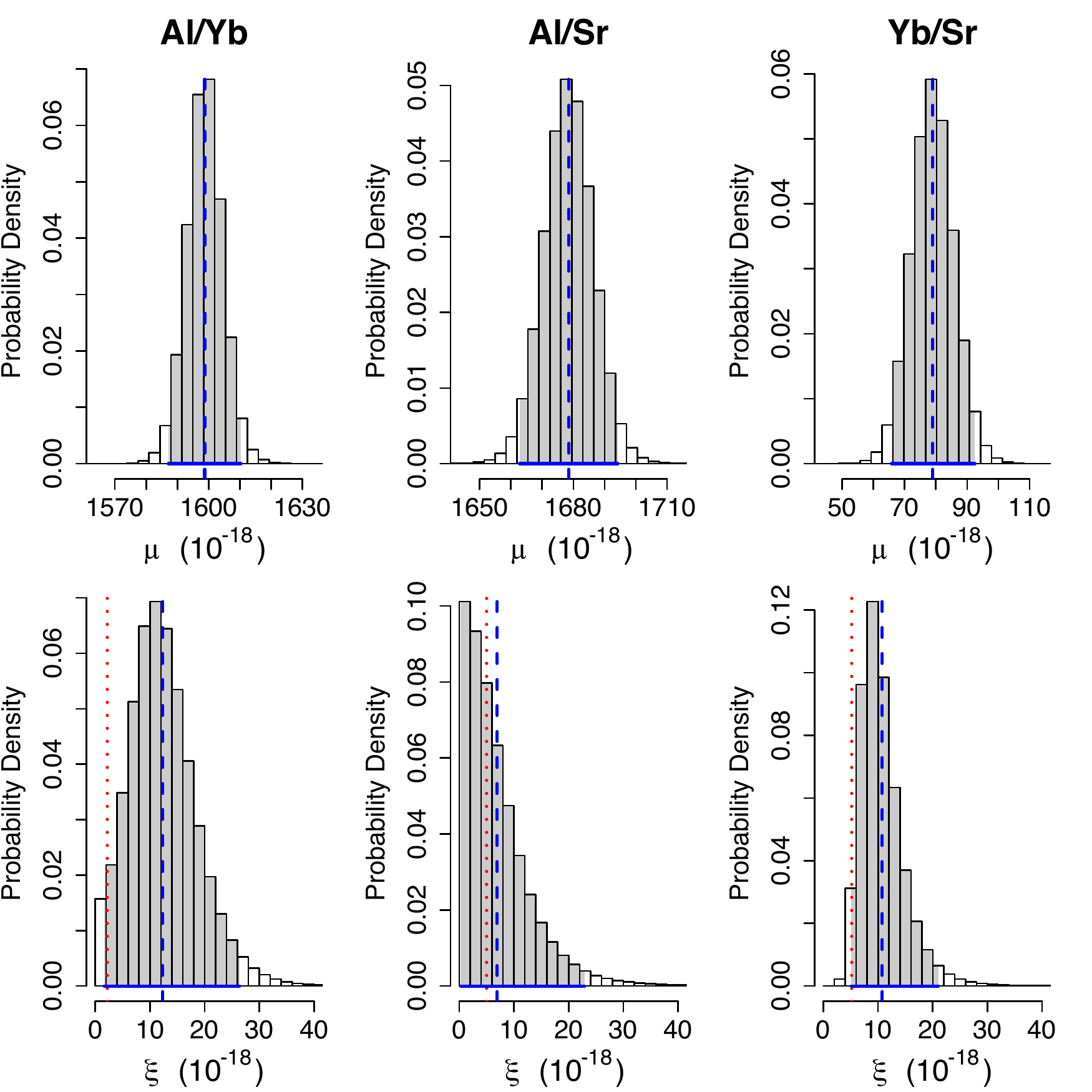}
\caption{Posterior distributions for the ratio values $\mu$, expressed here as a fractional offset from the current recommended CIPM values \cite{riehle2018}, and for the between-day variability $\xi$. The blue dashed lines denote our estimate for these parameters, the posterior mean. The shaded area and blue line on the bottom of each plot denote the 95\% credible interval for this estimate. For the $\xi$ distributions, these credible intervals are very wide. This is because there is not much information in the data to estimate this parameter; more comparison days would allow us to further constrain this distribution. The red dotted lines denote the static uncertainties due to systematic effects.}
\end{figure}

\newpage

\begin{figure}[t!]
\centering
  \includegraphics[width = 0.8\columnwidth]{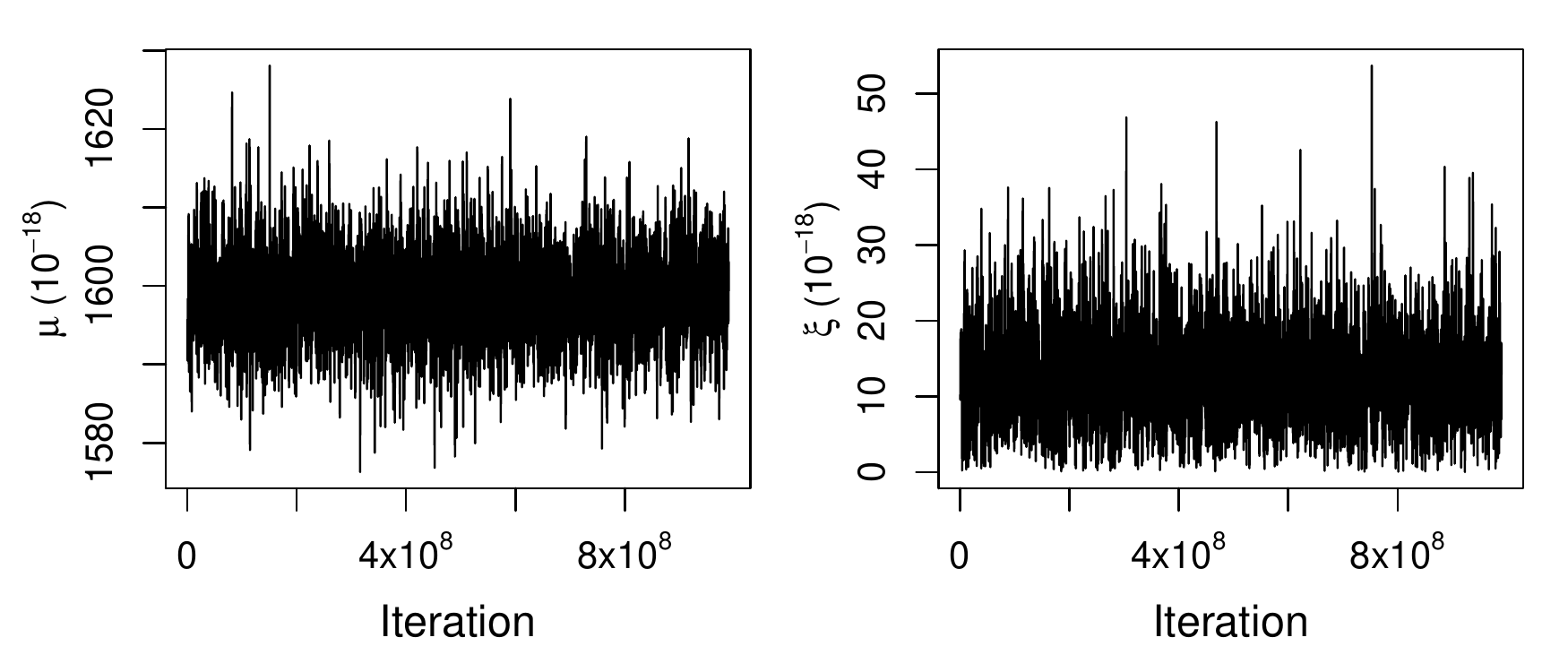}
\caption{Trace plots for the true Al$^+$/Yb ratio value $\mu$ (left), expressed here as a fractional offset from the current recommended CIPM values \cite{riehle2018}, and for the between-day variability $\xi$ (right). The x-axis is the MCMC iteration number and the y-axis is the value of the parameter. We plot only every 1000th sampled value. Trace plots are used as a convergence diagnostic for MCMC; these plots show that the chains are mixing well. A warning flag for poor mixing would be any obvious trend in the values over iteration. }
\end{figure}

\newpage
\renewcommand\thetable{S\arabic{table}}
\setcounter{table}{0}

\begin{table}[t!]
\small
\centering
\begin{tabular}{|r|c|c|c|}
\hline
\textbf{Parameter} & \textbf{$^{27}$Al$^{+}$} & \textbf{$^{171}$Yb} & \textbf{$^{87}$Sr}\\
\hline
Number of atoms        & 1    & 1000    &  2000 \\
Clock transition       & ${^1\!S_0}\leftrightarrow{^3\!P_0}$ & ${^1\!S_0}\leftrightarrow{^3\!P_0}$ & ${^1\!S_0}\leftrightarrow{^3\!P_0}$ \\
Clock wavelength [nm]        & 267    &  578   &   698  \\
Excited-state lifetime [s]       & 20.6(1.4)  &  20(2)  &    120(3)     \\
Nuclear spin      &  5/2 &  1/2  &  9/2  \\
Probe time [ms]       &  150  & 560  &  500  \\
Atom temperature [$\mu$K]      &  $\approx100$  & $\approx1$  &   $\approx2$  \\
Limiting systematic shifts      & \begin{tabular}{@{}c@{}}Micromotion\\time-dilation\end{tabular}  &  \begin{tabular}{@{}c@{}}Blackbody radiation,\\Lattice light\end{tabular}   &  Lattice light   \\
Servo time constant [s]      & 14    &  10  &   10   \\
Probe duty cycle [\%]        &  50   &  65   &   45 \\
\hline
\end{tabular}
\caption{Clock operating parameters.} 
\end{table}

\newpage

\begin{table}[t!]
\centering
\begin{tabular}{clrrr}
  \hline
Parameter & Ratio & Estimate & Uncertainty & 95\% CI \\
  \hline

& Al$^+$/Yb & 1598.7 & 5.9 & (1587.1, 1610.3) \\
$\mu$ ($10^{-18}$) & Al$^+$/Sr & 1678.5 & 8.0 & (1662.8, 1694.2) \\
& Yb/Sr & 79.0 & 6.8 & (65.8, 92.5) \\
  \hline
& Al$^+$/Yb &  12.3 & 6.2 & (1.6, 26.4) \\
$\xi$ ($10^{-18}$) & Al$^+$/Sr & 6.9 & 6.3 & (0.2, 22.9) \\
& Yb/Sr & 10.8 & 4.2 & (5.1, 21.1) \\
\hline
\end{tabular}
\caption{Results for $\mu$ ($10^{-18}$, expressed here as a fractional offset from the current recommended CIPM values \cite{riehle2018}) and $\xi$ ($10^{-18}$) based on 3960000 samples from the posterior distribution. The estimate is the mean of these samples, the uncertainty is the standard deviation, and the 95\% credible interval comes from the 95\% quantiles of this sample.}
\end{table}

\newpage

\begin{table}[t!]
\small
\centering
\begin{tabular}{|l|c|c|c|}
\hline
\textbf{Date (YYYY-MM-DD)} & \textbf{Al$^+$/Yb} ($x_i$) & \textbf{Al$^+$/Sr} ($x_j$) & $\textbf{Yb/Sr}$ ($x_k$) \\
\hline
2017-11-06 & 1.6067$\times 10^{-15}$ & n/a & n/a\\
2017-11-07 & 1.5879$\times 10^{-15}$ & n/a & n/a\\
2017-11-08 & 1.5819$\times 10^{-15}$ & n/a & n/a\\
2017-11-09 & 1.5891$\times 10^{-15}$ & n/a & n/a\\
2017-11-14 & 1.6667$\times 10^{-15}$ & n/a & n/a\\
2017-11-15 & 1.5729$\times 10^{-15}$ & n/a & n/a\\
2017-11-16 & 1.6106$\times 10^{-15}$ & n/a & n/a\\
2018-02-16 & 1.6323$\times 10^{-15}$ & n/a & 8.98$\times 10^{-17}$ \\
2018-02-28 & 1.6236$\times 10^{-15}$ & 1.6893$\times 10^{-15}$ & 7.94$\times 10^{-17}$ \\
2018-03-01 & 1.6019$\times 10^{-15}$ & 1.6766$\times 10^{-15}$ & 7.33$\times 10^{-17}$ \\
2018-03-05 & 1.6228$\times 10^{-15}$ & 1.6837$\times 10^{-15}$ & 8.37$\times 10^{-17}$ \\
2018-03-06 & 1.6156$\times 10^{-15}$ & 1.6854$\times 10^{-15}$ & 6.88$\times 10^{-17}$ \\
2018-03-09 & 1.5983$\times 10^{-15}$ & 1.6779$\times 10^{-15}$ & 8.77$\times 10^{-17}$ \\
2018-03-23 & n/a & n/a & 6.82$\times 10^{-17}$ \\
2018-03-28 & n/a & 1.6720$\times 10^{-15}$ & n/a\\
2018-04-03 & 1.6012$\times 10^{-15}$ & 1.6752$\times 10^{-15}$ & 7.59$\times 10^{-17}$ \\
2018-05-03 & 1.5747$\times 10^{-15}$ & n/a & n/a\\
2018-05-09 & 1.5802$\times 10^{-15}$ & n/a & n/a\\
2018-06-14 & 1.5795$\times 10^{-15}$ & 1.6784$\times 10^{-15}$ & 9.48$\times 10^{-17}$ \\
\hline
\end{tabular}
\caption{Daily measured offsets with respect to the ratios based on the current published CIPM recommended frequencies for each clock (Al$^+$: $\num{1121 015 393 207 857.3}$ Hz, Yb: $\num{518 295 836 590 863.6}$ Hz, and Sr: $\num{429 228 004 229 873.0}$) \cite{riehle2018}. These values correspond to the $x_{i}$, $x_{j}$, and $x_{k}$ used in the comprehensive model.}
\end{table}

\newpage

\begin{table}[t!]
\small
\centering
\begin{tabular}{|c|c|c|c|}
\hline
 & \textbf{Al$^+$} & \textbf{Yb}  & \textbf{Sr} \\
 & Total  & Total  & Total    \\
\hline
\diagbox{\textbf{Nominal date}\\ (YYYY-MM-DD)}{\\ \textbf{Model label}}   & $\mathbf{a}$    & $\mathbf{b}$    & $\mathbf{c}$   \\
\hline
2017-11-06   & 2.1$\times10^{-18}$    & 1.4$\times10^{-18}$    & n/a      \\
2017-11-07   & 2.1$\times10^{-18}$    & 1.4$\times10^{-18}$    & n/a      \\
2017-11-08   & 2.1$\times10^{-18}$    & 1.4$\times10^{-18}$    & n/a      \\
2017-11-09   & 2.1$\times10^{-18}$    & 1.4$\times10^{-18}$    & n/a      \\
2017-11-14   & 2.1$\times10^{-18}$    & 1.4$\times10^{-18}$    & n/a      \\
2017-11-15   & 2.1$\times10^{-18}$    & 1.4$\times10^{-18}$    & n/a      \\
2017-11-16   & 2.1$\times10^{-18}$    & 1.4$\times10^{-18}$    & n/a      \\
2018-02-16   & 2.1$\times10^{-18}$    & 1.4$\times10^{-18}$    & 4.91$\times10^{-18}$     \\
2018-02-28   & 2.1$\times10^{-18}$    & 1.4$\times10^{-18}$    & 4.82$\times10^{-18}$     \\
2018-03-01   & 2.1$\times10^{-18}$    & 1.4$\times10^{-18}$    & 4.74$\times10^{-18}$     \\
2018-03-05   & 2.1$\times10^{-18}$    & 1.4$\times10^{-18}$    & 4.53$\times10^{-18}$     \\
2018-03-06   & 2.1$\times10^{-18}$    & 1.4$\times10^{-18}$    & 4.84$\times10^{-18}$     \\
2018-03-09   & 2.1$\times10^{-18}$    & 1.4$\times10^{-18}$    & 4.84$\times10^{-18}$     \\
2018-03-23   & n/a    & 1.4$\times10^{-18}$    & 4.52$\times10^{-18}$     \\
2018-03-28   & 0.95$\times10^{-18}$   & n/a    & 4.53$\times10^{-18}$     \\
2018-04-03   & 0.95$\times10^{-18}$   & 1.4$\times10^{-18}$    & 4.75$\times10^{-18}$     \\
2018-05-03   & 0.95$\times10^{-18}$   & 1.4$\times10^{-18}$    & n/a      \\
2018-05-09   & 0.99$\times10^{-18}$   & 1.4$\times10^{-18}$    & n/a      \\
2018-06-14  & 0.99$\times10^{-18}$   & 1.4$\times10^{-18}$    & 6.78$\times10^{-18}$     \\
\hline
\end{tabular}
\caption{Daily uncertainties due to systematic effects, which correspond to $\mathbf{a}$, $\mathbf{b}$, and $\mathbf{c}$ in the comprehensive model. A more detailed discussion is given in Refs. \cite{mcgrew2018,brewer2019,bothwell2019} with additional considerations for this data offered in the text of the supplemental information. Entries labeled ``n/a'' have no valid data.}
\end{table}

\newpage

\begin{table}[t!]
\small
\centering
\begin{tabular}{|l|c|c|c|c|c|}
\hline
\textbf{Date} & \textbf{Al$^+$/Yb} & \textbf{Al$^+$/Sr} & $\textbf{Yb/Sr}$ & \textbf{Sr Density} & \textbf{Sr Density} \\
\hline
(YYYY-MM-DD) & $\sigma_{stab,i}$  & $\sigma_{stab,j}$ & $\sigma_{stab,k}$ & $\mu_{SrDen,\{j,k\}}$ & $\sigma_{SrDen,\{j,k\}}$ \\
\hline
2017-11-06 & $3.81\times 10^{-17}$ & n/a                   & n/a                   & n/a                   & n/a                    \\
2017-11-07 & $1.09\times 10^{-17}$ & n/a                   & n/a                   & n/a                   & n/a                    \\
2017-11-08 & $1.92\times 10^{-17}$ & n/a                   & n/a                   & n/a                   & n/a                    \\
2017-11-09 & $2.07\times 10^{-17}$ & n/a                   & n/a                   & n/a                   & n/a                    \\
2017-11-14 & $3.62\times 10^{-17}$ & n/a                   & n/a                   & n/a                   & n/a                    \\
2017-11-15 & $2.45\times 10^{-17}$ & n/a                   & n/a                   & n/a                   & n/a                    \\
2017-11-16 & $2.71\times 10^{-17}$ & n/a                   & n/a                   & n/a                   & n/a                    \\
2018-02-16 & $5.41\times 10^{-17}$ & n/a                   & $4.32\times 10^{-18}$ & $-5.02\times 10^{-17}$ & $2.48\times 10^{-18}$  \\
2018-02-28 & $1.99\times 10^{-17}$ & $3.62\times 10^{-17}$ & $8.80\times 10^{-18}$ & $-8.68\times 10^{-17}$ & $5.80\times 10^{-18}$  \\
2018-03-01 & $1.21\times 10^{-17}$ & $1.24\times 10^{-17}$ & $3.29\times 10^{-18}$ & $-3.07\times 10^{-17}$ & $4.24\times 10^{-18}$  \\
2018-03-05 & $1.57\times 10^{-17}$ & $2.37\times 10^{-17}$ & $3.59\times 10^{-18}$ & $-4.31\times 10^{-17}$ & $4.93\times 10^{-18}$  \\
2018-03-06 & $7.28\times 10^{-18}$ & $7.48\times 10^{-18}$ & $1.46\times 10^{-18}$ & $-7.96\times 10^{-17}$ & $3.83\times 10^{-18}$  \\
2018-03-09 & $3.24\times 10^{-17}$ & $3.28\times 10^{-17}$ & $2.27\times 10^{-18}$ & $-5.58\times 10^{-17}$ & $6.63\times 10^{-18}$  \\
2018-03-23 & n/a                   & n/a                   & $5.23\times 10^{-18}$ & $-4.73\times 10^{-17}$ & $4.44\times 10^{-18}$  \\
2018-03-28 & n/a                   & $9.46\times 10^{-18}$ & n/a                   & $-5.62\times 10^{-17}$ & $3.62\times 10^{-18}$  \\
2018-04-03 & $7.96\times 10^{-18}$ & $8.76\times 10^{-18}$ & $2.22\times 10^{-18}$ & $-7.34\times 10^{-17}$ & $6.48\times 10^{-18}$  \\
2018-05-03 & $1.06\times 10^{-17}$ & n/a                   & n/a                   & n/a                   & n/a                    \\
2018-05-09 & $1.91\times 10^{-17}$ & n/a                   & n/a                   & n/a                   & n/a                    \\
2018-06-14 & $2.00\times 10^{-17}$ & $2.20\times 10^{-17}$ & $1.88\times 10^{-18}$ & $-2.99\times 10^{-17}$ & $1.31\times 10^{-18}$  \\
\hline
\end{tabular}
\caption{Daily statistical uncertainties. These values are derived from fits to the daily Allan deviation of each ratio measurement and measurement uncertainties are quoted by extrapolation of the fit to the total measurement time using a white noise model. In addition to the statistics of the ratio measurement, both ratios involving Sr have an additional statistical measurement uncertainty due to the daily measurements of the $s$-wave density shift -- with average $\mu_{SrDen,k}$ and uncertainty $\sigma_{SrDen,k}$ -- in the Sr clock on each comparison day. These uncertainties make up the $\sigma_i$, $\sigma_j$, and $\sigma_k$ in the Bayesian model. For example, for the Yb/Sr ratio $\sigma_k^2 = \sigma_{stab,k}^2 + \sigma_{SrDen,k}^2$.}
\end{table}

\end{document}